%% file: template.tex
\documentclass{article}
\usepackage{arxiv}

\usepackage[utf8]{inputenc} 
\usepackage[T1]{fontenc}    
\usepackage{hyperref}       
\usepackage{url}            
\usepackage{booktabs}       
\usepackage{amsfonts}       
\usepackage{nicefrac}       
\usepackage{microtype}      
\usepackage{graphicx}
\usepackage{siunitx}
\usepackage{amsmath}
\usepackage{natbib}
\usepackage[color=pink]{todonotes}
\presetkeys{todonotes}{inline}{}
\setcitestyle{authoryear}
\usepackage{doi}
\usepackage{subfigure}
\usepackage{array}
\newcommand{\vect}[1]{\ensuremath{\boldsymbol{\mathbf{#1}}}}  
\newcommand{\matr}[1]{\ensuremath{\boldsymbol{\mathbf{#1}}}}  

\newcommand{\ncoarse}{\ensuremath{m_{\tilde h}}}  
\newcommand{\ncoarserho}{\ensuremath{m_{\tilde h,\rho}}}  
\newcommand{\ncoarsesigma}{\ensuremath{m_{\tilde h,\sigma}}}  
\newcommand{\nfine}{\ensuremath{m_{h}}}  
\newcommand{\T}{\ensuremath{^\mathrm T}}

\title{Flexible covariance structures on metric graphs}

\author{ \href{https://orcid.org/0009-0007-3180-3059}{\includegraphics[scale=0.06]{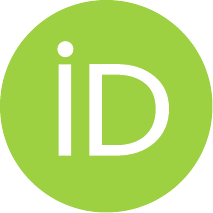}\hspace{1mm}Karina Lilleborge}\thanks{Corresponding author} \\
	Department of Mathematical Sciences\\
	Norwegian University of Science and Technology\\
	Trondheim, Norway \\
	\texttt{karina.lilleborge@ntnu.no} \\
	\And
	\href{https://orcid.org/0000-0003-4326-9029}{\includegraphics[scale=0.06]{figures/orcid.pdf}\hspace{1mm}Sara Martino} \\
	Department of Mathematical Sciences\\
	Norwegian University of Science and Technology\\
	Trondheim, Norway \\
	\texttt{sara.martino@ntnu.no} \\
	\AND
	\href{https://orcid.org/0000-0003-4995-2152}{\includegraphics[scale=0.06]{figures/orcid.pdf}\hspace{1mm}Geir-Arne Fuglstad} \\
	Department of Mathematical Sciences\\
	Norwegian University of Science and Technology\\
	Trondheim, Norway \\
	\texttt{geir-arne.fuglstad@ntnu.no} \\
}

\hypersetup{
	pdftitle={Flexible covariance structures on metric graphs},
	pdfsubject={stat.ME},
	pdfauthor={Karina Lilleborge, Sara Martino, Geir-Arne Fuglstad},
	pdfkeywords={Non-stationary, Gaussian random fields, SPDE approach, Metric graphs, Traffic modeling},
}

\begin{document}
\maketitle

\begin{abstract}
    Whittle-Matérn (WM) Gaussian random fields (GRFs) are defined as  solutions of stochastic partial differential equations (SPDEs) and provide a natural analog of  Matérn GRFs on non-Euclidean geometry where the Matérn covariance function is not valid. In particular, WM GRFs on metric graphs have been an active area of research motivated by  road and river networks where spatial dependence is more naturally described by  intrinsic distances in the network than by Euclidean distances. This family of GRFs is controlled by three parameters relating to marginal variance, spatial range, and smoothness, but can be extended to so-called generalized WM GRFs through spatially varying coefficients in the SPDE. Recent work has considered the use of spatially varying covariates, but the full possibilities of flexibility have not been considered. In this work, we introduce latent GRFs that describe the spatially varying coefficients of the SPDE. This flexible model is compared to less flexible models in a simulation study evaluating both the ability to estimate the covariance structure and predictive ability. An important focus is the number of observations and replications necessary to reliably recover the covariance structure. We find that the flexible model improves over less flexible models in the presence of sufficient data. We also demonstrate practical applicability on traffic counts in a part of Madrid, and observe major differences between in-sample and out-of-sample predictive abilities of the models compared.

\end{abstract}


\section{Introduction}\label{sec:introduction}
Gaussian random fields (GRFs) are an important tool for spatial modeling  and are popular for modeling real-life phenomena in a broad range of disciplines \citep{Diggle2007geostat,Gelfand2010, banerjee2026hierarchical}. In this paper, we consider the setting of a metric graph, which is a collection of edges, i.e., intervals, connected together in a graph structure. 
A key challenge in defining GRFs on non-Euclidean geometry such as metric graphs 
is to specify 
a useful family of valid covariance functions that can explain the spatial dependence structure and be estimated in practice.
Early work on streamflow networks focused on rivers described by tree structures that do not allow loops \citep{Hoef2006spatial,Hoef2010moving}.
However, more generally, one can combine geodesic distances or the resistance metric inspired by electrical network theory \citep{Anderes2020isotropic} with the Matérn covariance function, but one would, in general, be limited to smoothness
0.5 or smaller.

An appealing approach to overcome the complex geometry is to be inspired by the fact that a Matérn GRF $u(\cdot)$ on $\mathbb{R}^d$ can be defined as the stationary solution of a stochastic partial differential equation (SPDE)
\[
    (\kappa^2-\Delta)^{\alpha/2}(\tau u(\boldsymbol{s})) = \mathcal{W}(\boldsymbol{s}), \quad \boldsymbol{s}\in\mathbb{R}^d,
\]
where $\kappa>0, \tau >0$, $\alpha > d/2$, $\Delta$ is the Laplacian, and $\mathcal{W}(\cdot)$ is Gaussian white noise. A more detailed description will be given in Section \ref{sec:background}, but, intuitively, one can extend the Laplacian and the Gaussian white noise
to a metric graph $\Gamma$, and define a \emph{Whittle-Matérn} GRF (WM-GRF) $u(\cdot)$ on $\Gamma$ as the solution of
\begin{equation}
    (\kappa^2-\Delta)^{\alpha/2}(\tau u(\boldsymbol{s})) = \mathcal{W}(\boldsymbol{s}), \quad \boldsymbol{s}\in \Gamma, \label{eq:simpleSPDE}
\end{equation}
for $\alpha > 1/2$.

This provides a natural analog of the Matérn covariance structure on
a metric graph, with three parameters governing marginal variance, spatial range, and smoothness. However, the resulting covariance function is not, in general, a Matérn covariance function, which is generally not possible on a metric graph.
Instead the resulting process exhibits a behavior analogous to that of a Matérn GRF along the edges of the graph, while allowing for smoothness parameters larger than 0.5.
This extension of the ``SPDE approach'' \citet{Lindgren2011SPDE} to metric graphs is an active area of research \citep{Borovitskiy2021MaternOnGraphs,SanzAlonso2022SPDEGraph,Bolin2023statinfmetricgraph, Bolin2024GWMmetricgraph,Bolin2025loggausscox,Lilleborge2026}. 

\citet{Borovitskiy2021MaternOnGraphs} and \citet{SanzAlonso2022SPDEGraph} define a SPDE based stochastic model that is only defined at the vertices of the graph. 
Our interest, instead, is in spatial processes defined over the entire metric graph, that is, random functions $u: \Gamma \rightarrow \mathbb R$ whose realizations are defined both along the edges and at the vertices. We therefore follow the framework developed by  \citet{Bolin2025nonstat} who recently considered an extension of SPDE \eqref{eq:simpleSPDE} to spatially varying coefficients, $\kappa(\cdot)$ and $\tau(\cdot)$, under the name \emph{generalized Whittle-Matérn GRFs} (GWM-GRFs). Their work addresses theoretical properties of the model, introduces a finite element method (FEM), and studies numerical convergence. Their practical example considers a setting where repeated realizations can be split into two sets of data: (1) used to construct a spatial covariate and (2) used to estimate the model using the constructed covariate. In traffic applications; however, the observed road characteristics may not adequately explain local variation in traffic patterns \citep{LOWRY201498,PULUGURTHA2021103071}. While some information, such as road class or speed limits, may be available, many important determinants—including signal timing, traffic management strategies, temporary restrictions, driver behaviour, and the combined effects of local road geometry—are either unavailable, inconsistently recorded \citep{BarringtonLeigh2017, Haklay2010}, or difficult to represent through spatial covariates. This motivates more flexible models that do not require the covariance structure to be specified through observed covariates but instead allow it to vary locally.

We aim to consider the setting where the coefficients of the SPDE are themselves modeled as GRFs to allow more general variation at the same time as introducing penalization in the spirit of work such as \citet{Fuglstad2015doesnonstat}. This introduces a higher risk of overfitting the covariance structure, and it is important to assess how much data is needed to reliably fit such models. Further, we compare this more flexible approach to the standard WM-GRF and the data-splitting approach for the GWM-GRF. The goal is to provide guidance on model selection to practitioners working in settings with different data availability.
We focus both on the ability to determine the true covariance structure
and the predictive power at unobserved locations. The former is assessed by comparing the estimated and true spatially varying coefficients of the SPDE using root mean square error (RMSE). The latter is assessed through RMSE, mean absolute error (MAE), and continuous ranked probability score (CRPS) comparing the predictive distribution to the true signal in the simulation study and to unobserved values in the application.

The paper is organized as follows. Section~\ref{sec:background} provides a brief introduction to metric graphs, WM-GRFs and GWM-GRFs. Section~\ref{sec:model} describes the general setup for the Bayesian hierarchical models and the inference schemes. Then we describe the simulation study in Section~\ref{sec:simstudy}. In this study, we focus on spatial coverage and temporal resolution, and on the conditions under which models of different level of flexibility perform well. Lastly, in Section \ref{sec:case}, we present a case study on traffic intensity from Madrid, and end with discussion
in Section \ref{sec:discussion}.

\section{Gaussian Whittle-Mat\'ern fields on metric graphs}
\label{sec:background}
\subsection{Metric graphs}\label{ssec:metric_graph}
We refer to a metric graph $\Gamma$ as the pair $\Gamma=(\mathcal V,\mathcal E)$, where $\mathcal V$ is a set of vertices and $\mathcal E$ a set of edges. 
The edges $e\in\mathcal E$ are line segments that are attached in vertices $v \in \mathcal V$. We assume that the graph is connected, i.e., there exists a path between any pair of locations on the graph. A location on the graph is represented by the pair $\vect s=(e, d)\in \Gamma$, where $e\in\mathcal E$ identifies the edge and $d\in[0,1]$ gives the normalized position (parametrized by arc length) along edge $e$. 
We write $\underline e=(e,0)$ and $\overline e=(e,1)$ for the start and end points of edge $e$, respectively. Since vertices may be shared by multiple edges, a single location can admit multiple coordinate representations. For example, if the end vertex of $e_2$ coincides with the start vertex of $e_1$, then $\underline{e_1}=\overline{e_2}$, and both coordinate pairs represent the same spatial location.

The order of a vertex refers to the number of edges connected at the vertex. Note that there are multiple representations of a metric graph where vertices of order 2 can be added and/or removed, and the metric graph remains the same, while the minimal set of edges and vertices that represent the same geometry, is referred to as the minimal graph. 
Let $\lvert \mathcal V\rvert$ denote the number of vertices in the metric graph. Vertices of degree 1 are called terminal vertices, and vertices of order 2 or greater are internal vertices. 

A key component in extending the ``SPDE approach'' to metric graphs is to define a Laplacian, $\Delta_\Gamma$, on the metric graph. In our context, it is enough to define it almost everywhere, and we define it as a normal second-order derivative with respect to arc length on the internal points of the edges. However, under this definition, there is no unique inverse, and thus no unique solution to SPDE \eqref{eq:simpleSPDE}. Heuristically, this can be resolved by choosing that $\Delta_\Gamma^{-1} f$ for $f\in L_2(\Gamma)$ should be a function that is continuous at all vertices and where the directional derivatives sum to zero at all vertexes. \citet{Bolin2024GWMmetricgraph} gives a mathematical description of this Kirchhoff-Laplacian.

\subsection{SPDEs on metric graphs}\label{ssec:spde_graph}
We give a overview of model definition, parameterization, discretization, and computations in this section, but refer the reader to \cite{Bolin2024GWMmetricgraph, Bolin2023statinfmetricgraph,Bolin2025nonstat} for technical details and more formal conditions on the coefficients for WM-GRFs and GWM-GRFs, and \citet{Lilleborge2026} for a more practical introduction for WM-GRFs. This section focuses on known material described in an accessible way, and Section \ref{ssec:prop_model} focuses on the novel contributions in this paper.

Gaussian white noise, $\mathcal{W}(\cdot)$, extends in a straightforward way to a metric graph $\Gamma$ by measuring the sizes of sets using arclength, and the GWM-GRFs are defined
as solutions to
\begin{equation}\label{eq:spde_general}
	(\kappa(\vect s)^2-\Delta_\Gamma)^{\alpha/2}(\tau(\vect s) u(\vect s)) = \mathcal W(\vect s), \quad \vect s \in \Gamma,
\end{equation}
where $\Delta_\Gamma$ is the Kirchhoff-Laplacian, and $\kappa(\cdot)>\kappa_0>0$ and $\tau(\cdot)>\tau_0>0$ are real functions on $\Gamma$.  Note that $\mathcal W(\cdot)$ is an abuse of notation, as point-wise evaluation of the noise process is not defined.  The parameter $\alpha>1/2$ is assumed to be fixed and determines the smoothness of the solution 
$u(\cdot)$. At terminal vertices (i.e., vertices of order 1),  solving with the Kirchhoff-Laplacian implicitly imposes zero Neumann boundary conditions.


As discussed in the introduction, WM-GRF is not an approximation of a Matérn GRF on the metric graph, as the Matérn covariance function is, in general, not valid when geodesic distances are used. The covariance structure must be understood as
a generalization to the non-Euclidean geometry that locally behaves similarly as the Matérn, in the sense that it satisfies the same SPDE as it would in 1D together with vertex conditions to combine the different 1D segments together.
Unlike a Matérn GRF in the usual Euclidean case, a WM-GRF does not have a constant marginal variance $\sigma^2(\kappa, \tau, \alpha)$ and a constant practical correlation range $\rho(\kappa, \alpha)$.  A similar effect is found for bounded Euclidean domains where the boundary conditions affect the covariance structure in the vicinity of the boundary. 
This motivates us to avoid the imprecise terms stationary and non-stationary, and use WM-GRF for constant coefficients and GWM-GRF for spatially-varying coefficients.

Inspired by \citet{Lindgren2011SPDE}, \citet{Bolin2025nonstat} propose using log-regressions for the spatially varying coefficients,
\begin{align}\label{eq:logreg_k_t}
	\log\kappa(\vect s) = \sum^{p^\kappa}_{j=1} b_j^\kappa\xi^{\kappa}_j(\vect s), \quad\quad \log\tau(\vect s) = \sum^{p^\tau}_{j=1} b_j^\tau\xi^{\tau}_j(\vect s), \quad \vect s \in\Gamma,
\end{align}
where $p^\kappa,p^\tau\in\mathbb{N}$ denotes the numbers of basis functions, $\{\xi_1^\kappa(\cdot), \ldots,\xi_{p^\kappa}^\kappa(\cdot)\}$ and $\{\xi_1^\tau(\cdot),\ldots,\xi_{p^\tau}^\tau(\cdot)\}$ are  known real-valued basis functions, and the coefficients $b_1^\kappa,\ldots,b_{p^\kappa}^\kappa\in\mathbb{R}$ and $b_1^\tau,\ldots,b_{p^\tau}^\tau \in\mathbb{R}$ are model parameters.
For $p^\kappa=p^\tau=1$ and $\xi^\kappa(\cdot)=\xi^\tau(\cdot)\equiv 1$, this gives a WM-GRF. For small $p^\kappa$ and $p^\tau$, and  $\xi^\kappa(\cdot)$ and $\xi^\tau(\cdot)$ defined as covariates, this gives GWM-GRFs where covariance structure is described by a few spatial covariates. Finally for large $p^\kappa$ and $p^\tau$, this gives GWM-GRFs with a very flexible covariance structure. The two former models have been detailed in existing literature, while the latter is novel and is explained in Section \ref{ssec:prop_model}.

Computations with WM-GRFs on metric graphs are based on extending the finite element method (FEM) approach of \citet{Lindgren2011SPDE} from Euclidean domains to metric graphs. The FEM representation leads to sparse element matrices and, consequently, sparse precision matrices, which provide the Markov properties exploited for efficient computation. The key idea is to use the fact that each edge is locally one-dimensional and to seek a weak solution to \eqref{eq:spde_general} that is compatible across the graph. In the simplest case, where the graph consists of a single edge connecting two terminal vertices, the solution coincides with the classical one-dimensional Matérn Gaussian random field (up to boundary effects). We fix $\alpha = 2$ for the rest of the paper, which gives a smoothness of $1.5$ so that GRF is once differentiable, everywhere except the vertices, and the derivative is a GRF with smoothness $0.5$. The conditions on
the inverse of $\Delta_\Gamma$ ensures that, for each vertex, the GRF $u(\cdot)$ is
continuous and that the directional derivatives sum to zero.

\begin{figure}[htb]
	\centering
	\includegraphics[width=0.5\linewidth]{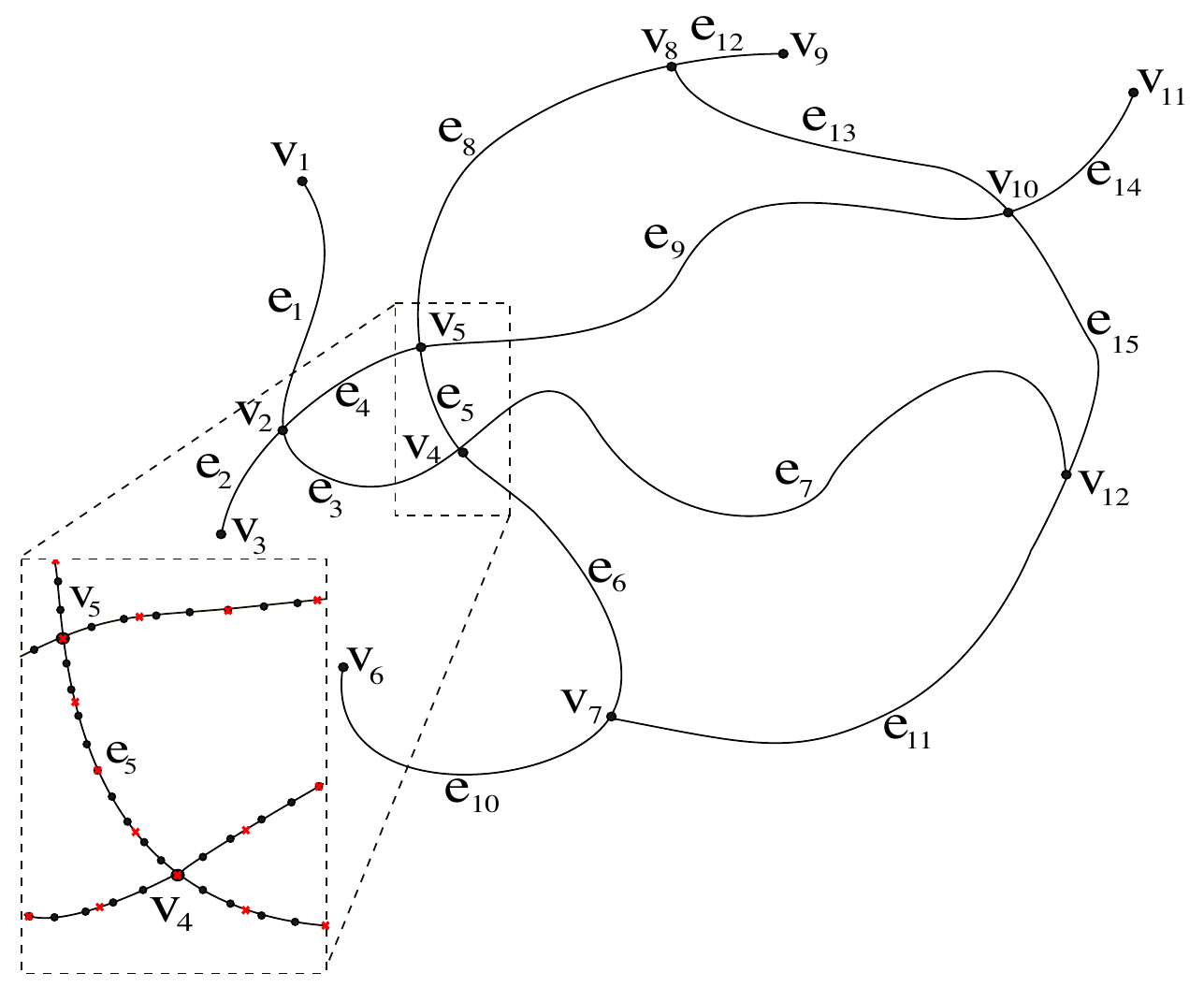}
	\caption{Illustration of a simple metric graph with 15 edges and 12 vertices. The zoomed panel to the left illustrates two different types of mesh vertices added with different maximal distance $h$ between mesh vertices. One mesh with shorter distance between vertices is shown in black dots ($\cdot$)) and another mesh with larger distance between vertices is shown with red crosses (\color{red}$\times$\color{black}).}
	\label{fig:graph_mesh_illustration}
\end{figure}

To apply FEM, we discretize the metric graph by constructing a mesh. 
This is done, by adding vertices to edges that are longer than a certain maximal mesh spacing $h>0$. For an edge with length larger than $h$, new vertices are inserted recursively until all resulting sub-edges have lengths no greater than $h$. The resulting metric graph, consisting of the original vertices together with the additional mesh vertices, is referred to as the mesh.
The mesh is ``the same'' metric graph as originally, but not the minimal description without vertices of order 2. 
Critically, \citet{Bolin2024GWMmetricgraph} show that adding a vertex, of order 2, to an edge does not change the solution $u(\cdot)$. Let $\mathcal V_h\supseteq\mathcal V$ be the set of vertices in the mesh and $\nfine=\lvert\mathcal V_h\rvert$. We refer to Figure~\ref{fig:graph_mesh_illustration} for an illustration of a simple metric graph with two meshes of different coarseness $h$.

\begin{figure}[htb]
	\centering
	\includegraphics[width=0.7\linewidth]{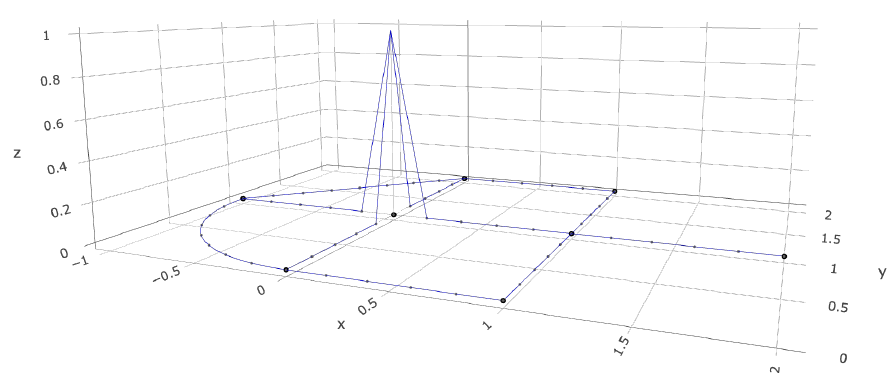}
	\caption{Piecewise linear basis function on a simple metric graph. This is a basis function for a so-called graph vertex.}
	\label{fig:basis_function_illustration}
\end{figure}

Given the  mesh with maximal spacing $h$, we  define the set of linear basis functions $\{\psi_{h,j}(\cdot)\}_{j=1}^{\nfine}$ as the collection of piecewise linear functions that satisfy $\psi_{h,j}(v_i)=1$ if and only if $j=i$ and zero otherwise for a graph or mesh vertex $ v_i\in\mathcal V_h$. Such set can be defined for any mesh on a graph $\Gamma$. The number of basis functions is equal to the number of vertices, $\nfine$, in the mesh. Figure~\ref{fig:basis_function_illustration} illustrates one basis function on a simple graph.

The FEM approach to obtain the weak solution to \eqref{eq:spde_general} on $\Gamma$ with spatially varying $\kappa(\vect s)$ and $\tau(\vect s)$ was introduced by \citet{Bolin2025nonstat}. In the following we provide a short introduction here, and refer the reader to \citet{Bolin2025nonstat} for full derivation. 
We approximate the solution $u(\cdot)$,  on a mesh with vertices $\mathcal V_h$, using the finite-dimensional representation\begin{equation*}
	u_h(\vect s) = \sum_{j=1}^{\nfine} w_j \psi_{h,j}(\vect s),\quad \vect s\in\Gamma,
\end{equation*}
where $\psi_{h,j}(\cdot)$ for $j=1,\ldots,\nfine$ are piece-wise linear basis functions and $\nfine=|\mathcal V_h|$. The weak formulation is obtained by requiring that,
\begin{equation*}
	\begin{bmatrix}
		\int_\Gamma (\kappa(\vect s)^2-\Delta)(\tau(\vect s) u_h(\vect s)) \psi_{h,1}(\vect s) ds \\
		\vdots \\
		\int_\Gamma (\kappa(\vect s)^2-\Delta)(\tau(\vect s) u_h(\vect s)) \psi_{h,\nfine}(\vect s) ds
	\end{bmatrix} \overset{\mathrm{d}}{=} \begin{bmatrix}
		\int_\Gamma \mathcal W(\vect s) \psi_{h,1}(\vect s) ds \\
		\vdots \\
		\int_\Gamma \mathcal W(\vect s) \psi_{h,\nfine}(\vect s) ds
	\end{bmatrix} 
\end{equation*}
By the divergence theorem, and assuming Neumann boundary conditions for $u_h(\cdot)$ and $\tau(\cdot)$ (the derivatives are zero at the boundary/terminal vertices), we end up with the following equality for each row
\begin{equation*}
	\int_\Gamma \kappa(\vect s)^2\tau(\vect s) u_h(\vect s) \psi_{h,k}(\vect s) ds +  \int_\Gamma \nabla(\tau(\vect s) u_h(\vect s)) \nabla \psi_{h,k}(\vect s) ds =^d \int_\Gamma \mathcal W(\vect s) \psi_{h,k}(\vect s) ds 
\end{equation*}
for all $k=1,\ldots,\nfine$. We can define element matrices $\matr C_h=[\langle \psi_{h,i},\psi_{h,j}\rangle]_{ij}$, $\matr C_h^{\kappa,\tau}=[\langle \tau\kappa^2 \psi_{h,i},\psi_{h,j} \rangle]_{ij}$ and $\matr G_h^{\tau}=[\langle \nabla(\tau \psi_{h,i}), \nabla\psi_{h,j}\rangle]_{ij}$, where $\langle f,g\rangle$ is integration over all of $\Gamma$, which can be understood as piecewise integration over the collection of edges. In practice, one cannot evaluate the integrals for all elements in the matrices $\matr C^{\kappa,\tau}_h$ and $\matr G_h^{\tau}$. Instead one approximates $C^{\kappa,\tau}$ by combining a lumped mass matrix $\matr C_h$ and diagonal matrices of $\kappa(\cdot)$ and $\tau(\cdot)$ evaluated in vertices $\mathcal V_h$.
We further obtain the precision matrix for the weights $\vect w=[w_1\:\ldots\:w_{\nfine}]\T$, given by
\begin{equation}
	\matr Q_{h,2} = (\matr C_h^{\kappa,\tau} + \matr G_h^{\tau})\T\matr C_h^{-1}(\matr C_h^{\kappa,\tau} + \matr G_h^{\tau}).
\end{equation}

Note that, for spatially constant coefficients in \eqref{eq:spde_general}, \ $\kappa(\vect s)=\kappa$ and $\tau(\vect s)=\tau$ for all $\vect s\in\Gamma$, the precision matrix simplifies to
\begin{equation}\label{eq:prec_u_stat}
	\matr Q_{h,1} =\tau^2(\kappa^2\matr C_h + \matr G_h)\T \matr C_h^{-1}(\kappa^2\matr C_h + \matr G_h),
\end{equation}
where $\matr G_h=[\langle \nabla(\psi_{h,i}), \nabla\psi_{h,j}\rangle]_{ij}$ and still $\matr C_h=[\langle \psi_{h,i},\psi_{h,j}\rangle]_{ij}$. 

A common way to specify the spatially varying $\kappa(\vect s)$ and $\tau(\vect s)$, introduced in \citet{Ingebrigtsen2015} for Euclidean domains and later tested for metric graphs by \citet{Bolin2025nonstat}, is to parameterize the coefficient fields through spatial covariates. Specifically:
\begin{align}\label{eq:cov_k_t}
	\log\kappa(\vect s) = b_1^\kappa+({\vect b_2^\kappa})\T \vect z^\kappa(\vect s), \quad\quad \log\tau(\vect s) = b_1^\tau +({\vect b_2^\tau })\T\vect z^\tau(\vect s), \quad \vect s \in \Gamma,
\end{align}
where $\vect z^\kappa(\cdot)$ and $\vect z^\tau(\cdot)$ are known covariates defined over the entire graph. 
The choice of covariates is important, as appropriately selected covariates can provide an interpretable representation of the sources of non-stationarity. For example, altitude covariates can be used to explain spatial variation in rainfall patterns, as demonstrated by \citet{Ingebrigtsen2015}. When covariate information is only available at a finite set of locations rather than throughout the entire domain, \citet{Bolin2025nonstat} propose covariate smoothing methods to obtain spatially continuous coefficient fields.

\subsection{Proposed model for GWMF}\label{ssec:prop_model}
Our goal is to introduce an alternative and more flexible way to define covariance structures for the field defined in \eqref{eq:logreg_k_t}. 
In the following, inspired by \citet{Lindgren2011SPDE}, we reparameterize our model as:
\begin{equation}\label{eq:param_transformation}
	\rho(\vect s)=2\sqrt{3}/\kappa(\vect s) \quad\text{and}\quad \sigma(\vect s)=\frac{1}{2[\kappa(\vect s)]^{3/2}\tau(\vect s)}, \quad\vect s\in \Gamma,
\end{equation}
where we have assumed smoothness $\alpha=2$ ($d=1$, $\nu=1.5$). 
In the Euclidean case, $\rho$ and $\sigma$ can be interpreted as the practical range and the marginal variance of the Matèrn field.
In the metric graph setting, $\rho(\cdot)$ and $\sigma(\cdot)$ should be interpreted as \textit{approximate} correlation range and marginal standard deviation parameters due to the influence of graph topology and vertex conditions on the covariance structure.
Note that the transformation is linear in log scale, hence inference can equivalently be performed in terms of either parameterization.


To model spatially varying coefficients $\rho(\cdot)$ and $\sigma(\cdot)$, we introduce two coefficient meshes on the graph with maximal edge length $\tilde h_\rho$ and $\tilde h_\sigma$ respectively. Denote the corresponding sets of mesh vertices by $\mathcal V_{\tilde h_\rho}$ and $\mathcal V_{\tilde h_\sigma}$ and the numbers of vertices by $\ncoarserho$ and $\ncoarsesigma$ respectively. 
 
The spatially varying coefficient fields are then represented as
\begin{align}\label{eq:basis_rho_mvar}
	\log\rho(\vect s) = \sum^{\ncoarserho}_{j=1} b_j^\rho\varphi^\rho_{\tilde h_\rho,j}(\vect s) \quad\text{and}\quad \log\sigma(\vect s) = \sum^{\ncoarsesigma}_{j=1} b_j^{\sigma}\varphi^\sigma_{\tilde h_\sigma,j}(\vect s),
\end{align}
where $\{b_j^a\}_{j=1}^{\ncoarse}$ for $\ncoarse=\ncoarserho,\ncoarsesigma$ are the weights associated with the two coefficient fields.

To use this flexible parameterization of $\rho(\cdot)$ and $\sigma(\cdot)$ appropriate priors or penalization terms must be assigned to the coefficients $b_j^\rho$, $j=1,\ldots,\ncoarserho$, and $b_j^\sigma$, $j=1,\ldots,\ncoarsesigma$ to avoid over-fitting. Highly flexible models may fit well to the data, but behave poorly when used for predictions either in unobserved locations or of future observations.
On the other side, overly restrictive models may fail to capture more advanced spatial dependencies.
Therefore, a balance between model flexibility and regularization is required.
From here on, we will only consider $\mathcal V_{\tilde h_\rho}=\mathcal V_{\tilde h_\sigma}$ and $\ncoarserho=\ncoarsesigma=\ncoarse$ and the maximal distance between mesh vertices be $\tilde h$. The sets of basis functions become a shared set $\{\varphi_{\tilde{h},j}\}_{j=1}^{\ncoarse}$.

We force some relatedness between the coefficients $\{b^\rho_j\}_j$ and between $\{b^\sigma_j\}_j$ by using the same approach suggested by \citet{Fuglstad2015nonstat}. Let $p(\cdot)$ be a GRF defined on a graph $\Gamma$ (it can be either $\log\rho(\vect s)$ or $\log\sigma(\vect s)$), which can be represented as the solution to an SPDE. First, let $p(\cdot)$ be described as follows
\begin{equation}\label{eq:general_f}
	p(\vect s) = p_0+\sum ^{\ncoarse}_{j=1} b^p_j \varphi_{\tilde{h},j} (\vect s), \quad \vect s\in \Gamma,
\end{equation}
where $p_0$ refers to the level and $\varphi_{\tilde{h},j}(\cdot)$, $j=1,\ldots,\ncoarse$, are piecewise linear basis functions on the mesh with maximal spacing $\tilde h$. To ensure identifiability of $p_0$ we add a sum-to-zero constraint on $\{b^p_j\}_{j=1}^{\ncoarse}$. We choose $p(\vect s)$ as the weak solution to a SPDE of the form
\begin{equation}\label{eq:spde_f}
		- \Delta_\Gamma (p(\vect s)) = \mathcal W_p(\vect s), \quad \vect s \in\Gamma,
\end{equation}
where $\mathcal W_p(\vect s)$ is Gaussian white noise without the 0-frequency. See Appendix \ref{app:A} for more details on this. 
The precision matrix of the vector $[b^p_1\:\ldots \: b^p_{\ncoarse}]\T$ is given by
\begin{equation*}
	\matr G_{\tilde h}\T \matr C_{\tilde h} ^{-1}\matr G_{\tilde h},
\end{equation*}
where $\matr C_{\tilde h}$ and $\matr G_{\tilde h}$ are element matrices related to the discretization of $\Gamma$ with maximal distance between vertices $\tilde h$ and the basis $\{\varphi_i (\vect s)\}_{i=1}^{\ncoarse}$. These matrices are easy to compute, and are only non-zero in neighboring elements where basis functions are both non-zero.
This matrix has null vector $\mathbf 1$, and is not invertible and therefore not a valid covariance structure. To handle this, we suggest a modified precision matrix
\begin{equation}\label{eq:mod_Q}
	\matr{Q}_{\tilde h,3} (\vect \tau)= \tau_1\matr G_{\tilde h} \matr C_{\tilde h} ^{-1} \matr G_{\tilde h} + \tau_2\vect 1 \vect 1\T
\end{equation}
where $\tau_1$ and $\tau_2$ are penalty parameters, ($\vect \tau= [\tau_1\: \tau_2]\T$) to obtain penalization for the complexity of the fields, $\log\rho(\cdot)$ and $\log\sigma(\cdot)$.
The interpretation of $\tau_1$ is that increasing $\tau_1$, also increases the relatedness between neighboring elements, and increasing $\tau_2$ means that we force the field to zero (we increase precision for each coefficient \textit{and} the relatedness between them). In practice, the value of $\tau_1$ is the parameter that needs to be tuned for, while minor changes in $\tau_2$ does little to the prior/penalization of complex models. 

We suggest to tune the penalization parameter $\tau_1$ by Monte Carlo simulation with the known distribution of $\vect b^f=[b^f_1\:\ldots\:b^f_{\ncoarse}]\T$. By constructing a mesh on the graph with $\ncoarse$ vertices, we can draw samples from $\mathcal N_{\ncoarse}(\vect 0, \matr Q_{\tilde h,3}^{-1})$, and compute the 95\%-quantile for each mesh location, and further evaluate if the range of these quantiles is sufficiently large compared to how much $\log\rho(\cdot)$ and $\log\sigma(\cdot)$ vary in space in our prior knowledge about the log-transformed fields. 

Prior information about the practical range can be specified through the implied log-Gaussian distribution of $\rho(\vect s)$. One must have some prior assumption on the mean and variance of $\rho(\vect s)$. Then, one can find the corresponding prior mean and variance of $\log\rho(\vect s)$ through
$$
\mu_{\log\rho} = \log\left(\frac{\mu_\rho}{\sqrt{\frac{\sigma_\rho^2}{\mu_\rho^2}+1}}\right)\quad\quad \sigma_{\log\rho}^2 = \log\left(\frac{\sigma_\rho^2}{\mu_\rho^2}+1\right)
$$
where $\log(\cdot)$ is the natural logarithm, $\mu_\rho$ is the prior mean for $\rho(\cdot)$ and $\sigma^2_\rho$ is the prior variance. Having prior knowledge of $\mu_{\log\rho}$ and $\sigma_{\log\rho}^2$ can then be imposed in tuning for $\tau_1$ by ensuring enough variability in the 95\%-quantiles.

In practice, we solve \eqref{eq:spde_general} using FEM and we choose mesh distance $h$ such that the mesh used for FEM for the latent field $u(\cdot)$ is fine enough to obtain sufficiently accurate results, and the mesh distance $\tilde h$ used for basis functions in the covariance structure is reasonable with respect to estimating the underlying field from the data.
Prior specification is closely tied to the mesh resolutions used in the approximations.
The coarse mesh used for $\log\rho(\cdot)$ and $\log\sigma(\cdot)$ controls the spatial scale on which the covariance structure is allowed to vary, and therefore also the number of parameters that must be estimated. It should be chosen fine enough to represent the expected non-stationarity, but coarse enough that the resulting coefficient fields remain identifiable from the available data.
The fine mesh used in the FEM approximation of the latent field must also be compatible with the prior range values allowed by the model. In particular, the prior for $\rho(\cdot)$ should not put substantial mass on correlation ranges that are shorter than the numerical resolution of the fine mesh. We therefore choose the prior mean and variance for $\log\rho(\cdot)$ together with the fine mesh size $h$, so that the mass of prior distribution for $\rho(\cdot)$ remains above the mesh scale. This provides a practical link between prior specification and numerical accuracy.

\section{Model}\label{sec:model}
\subsection{Model specification}
We consider $R$ sets of observations $y^r(\vect s_k)$ at spatial locations $\vect s_k\in\mathcal S^r\subset \Gamma$ for $r=1,\ldots, R$. Let $N_r=\lvert \mathcal S^r\rvert$ indicate the number of observations of replicate $r$. We assume the observations are conditionally independent given some underlying process $\eta^r(\cdot)$ and parameters,
\begin{equation}\label{eq:obs}
	y^r(\vect s^r_k) = \eta^r(\vect s^r_k) + \varepsilon^r_k,\quad \varepsilon^r_k|\sigma_\mathrm{N}^2\sim \mathcal N(0,\sigma_\mathrm{N}^2), \: \vect s^r_k\in\mathcal S^r, \: r=1,\ldots, R,
\end{equation}
where $\eta^r(\cdot)$ is a sum of spatially varying covariates and a (G)WM-GRF,
\begin{equation}\label{eq:latent}
	\eta^r(\vect s) =\vect x (\vect s)\T \vect \beta +  u^r(\vect s),\quad \vect s \in\Gamma,
\end{equation}
where $\vect x(\cdot)=[x_1(\cdot)\:\ldots\: x_p(\cdot)]\T$ contains spatial covariates describing the mean structure, $\vect \beta$ is a $p\times 1$ vector of regression coefficients which includes an intercept, and $u^r(\cdot)$, $r=1,\ldots,R$, are independent replicates of (G)WM-GRFs defined through the SPDE representation in \eqref{eq:spde_general}.

For inference, the model is represented using a finite element discretization of the graph $\Gamma$. Specifically, we construct a mesh with spacing $h$ as described in Section~\ref{ssec:spde_graph} and let $\nfine$ indicate the number of vertices in this mesh. The latent fields $u^r(\cdot)$ are then represented using the corresponding finite element basis functions, with weights  $\vect w^r = [w^r_1 \:\ldots\: w^r_{\nfine}]^\mathrm{T}$ for $r=1,\ldots, R$. 
In addition to the fine mesh used for representation of the (G)WM-GRF $u^r(\cdot)$, the proposed model introduces a second, coarser mesh with spacing $\tilde{h}$ and
$\ncoarse$ vertices for modeling the spatial variation in the covariance parameters. This mesh defines $\ncoarse$ basis functions used to represent the coefficient fields $\log\rho(\cdot)$ and $\log\sigma(\cdot)$, with corresponding weights  $\{b_j^\rho\}_j$ and $\{b^\sigma_j\}$. 
The use of a coarser mesh for the covariance parameters reduces computational complexity, since increasing the number of basis functions quickly increases the dimensionality of the model. In addition, the spatial resolution of the covariance fields should be informed by the available data density, as variations in the covariance structure at scales much smaller than the observation spacing are difficult to identify from the data.


We define the vector with all stacked observations:
$$
\vect Y =\left[({\vect y^1})\T\:\ldots\:({\vect y^R})\T\right]\T, 
$$ 
where $\vect y^r=[y^r(\vect s^r_1)\:\ldots\: y^r(\vect s^r_{N_r})]\T$ for $r=1,\ldots, R$. Let $n=\sum^R_{r=1}N_r$ be the total number of observations.
Furthermore, let 
$$
\matr X=\left[{\matr x^1}\:\ldots\:{\matr x^R}\right]\T \quad\text{ and }\quad \vect W = \left[ ({\vect w^1})\T\:\ldots\: ({\vect w^R})\T \right]\T,
$$
where $\matr X$ is a $p\times n$ matrix built from the $R$ matrices of dimension $p\times N_r$, $\matr x^r=[\vect x(\vect s^r_1)\:\ldots\:\vect x(\vect s^r_{N_r})]$, $r=1,\ldots,R$, containing information about spatial covariates,  
and $\vect W$ is a $(\nfine\cdot R)\times 1$ vector containing all weights for all $R$ replicates. 

We define $\matr A_h$ as the projection matrix mapping the finite element weights  to the observation locations.  Specifically, $\mathbf A_h$ is an $n\times (\nfine \cdot R)$ matrix that evaluates the latent fields represented by the weight vectors $\mathbf w^r$ at the observed locations in $\mathcal S^r$, for $r=1,\ldots,R$. 
The hierarchical model is
\begin{equation}\label{eq:hierarchical_model}
	\begin{aligned}
		\vect Y \mid \vect W,\vect \beta, \sigma_\mathrm{N}^2 &\sim \mathcal N_{n}(\matr X\T \vect \beta +\matr A_h\vect W, \sigma_\mathrm N^2 \matr I) \\
		\vect w^r \mid \vect \theta &\sim \mathcal N_{\nfine}(\vect 0, \matr Q_{h,q}(\vect \theta)^{-1}),\quad r=1,\ldots,R, \\
		1/\sigma_\mathrm N^2 &\sim \text{Gamma}(\alpha_\sigma, \beta_\sigma) \\
		\vect\beta &\sim \mathcal N_p(\vect 0, \matr Q_\beta^{-1}) \\
		\vect \theta &\sim \mathcal N(\vect \mu_\theta, \matr Q_{\theta}^{-1}) 
	\end{aligned}
\end{equation}
where $\vect w^r$ are the weights associated to replicate $r$ and basis $\{\psi_{h,j}\}_{j=1}^{\nfine}$. $\matr Q_{h,q}$ is the precision matrix of these weights where $q=1,2$ indicates if we consider a WM-GRF or GWM-GRF representation for $\eta^r(\cdot)$, $r=1,\ldots,R$. $\vect \theta$ contains the hyperparameters that enter the precision matrix $\matr Q_{h,q}(\vect \theta)$, $q=1,2$, and the prior precision $\matr Q_\theta$ can be the suggested $\ncoarse \times \ncoarse$ precision matrix $\matr Q_{\tilde h, 3}(\vect \tau)$ when we consider a GWM-GRF with the flexible parameterization described in \eqref{eq:basis_rho_mvar}. 
When we consider WM-GRFs with spatially constant coefficients, the precision matrix $\matr Q_\theta$ is only $2\times 2$, and GWM-GRFs with parameterization as in \eqref{eq:cov_k_t} we have additional rows and columns for each covariate included in $\vect z^\kappa(\cdot)$ and $\vect z^\tau(\cdot)$. 
To complete the model, we assign an inverse Gamma prior to the noise variance
$\sigma_\mathrm{N}^2$  with shape 1 and rate  $5\cdot 10^{-5}$. For $\boldsymbol\beta$, we use a Gaussian prior with zero mean. The precision is 0 for the intercept $\beta_0$ and  $10^{-3}$ for the remaining regression coefficients. 

\subsection{Inference and implementation details}
The number of parameters to be estimated is substantially increased when we consider the flexible proposed model, and we opt for an empirical Bayes procedure using the
maximum a posteriori estimates of the parameters, $\vect \theta$, based on $\hat {\vect \theta}|\matr Y$.
I.e., uncertainty in parameters are not propagated into the predictive distributions for the most flexible model.
The other two models are considered in a fully Bayesian framework, as the number of hyperparameters is low enough that the computational cost is not a limiting factor.

When we have obtained estimates $\hat {\vect \theta}$, we can make prediction in new locations in the domain $\mathcal S^r_{*}$ for replicate $r=1,\ldots, R_{*}$. 
Predictions and their uncertainties are obtained from linear transformations of the posterior distribution mean and variance of
$\vect\eta^r\mid \vect y^r,\vect\theta=\hat{\vect\theta}$ when we consider our proposed flexible model, while when we consider models with lower dimensional hyperparameter space, we can use a linear transformation of the posterior mean of $\vect\eta^r \mid \vect y^r$ and the posterior variance-covariance for uncertainty. Note that for the flexible proposed model, the uncertainties of estimates of $\vect\theta$ are not propagated in this step.

We use \texttt{MetricGraph} \citep{Bolin2023MetricGraph} to represent the metric graph and \texttt{rSPDE} \citep{Bolin2019rSPDE} to handle SPDE models on metric graphs represented as \texttt{MetricGraph}-objects. The \texttt{MetricGraph}-package handles representations of a metric graphs from spatial geometries from \texttt{sf} \citep{sf, sf1}, which can be obtained from \texttt{OpenStreetMap} \citep{osm}, and \texttt{rSPDE} writes model classes that \texttt{R-INLA} \citep{Rue2009INLA} can interpret for inference. Specifically, we use the wrapper library \texttt{inlabru} \citep{Lindgren2024inlabru} for simple model specification. The Empirical Bayes strategy is implemented in \texttt{R-INLA} and can be set using \texttt{int.strategy="eb"}. For the models with few hyperparameters, we perform a full Bayesian analysis with \texttt{R-INLA}, but when the dimension of the hyperparameter space becomes large for the most flexible model, this becomes infeasible, and we use empirical Bayes. 

\section{Simulation study}\label{sec:simstudy}
\subsection{Motivation and goal}
We conduct a simulation study to assess parameter identifiability and predictive performance for the models introduced in Section~\ref{sec:model} under controlled settings with varying levels of spatial and temporal coverage.
More specifically, we  compare three models for describing the spatial components $\{u^1(\cdot), \ldots, u^R(\cdot)\}$. As a baseline,  we consider a model  with spatially constant covariance parameters $\kappa(\cdot)$ and $\tau(\cdot)$, which we refer to as the \textsc{WMF} model. We then consider a second model, in which  
$\kappa(\cdot)$ and $\tau(\cdot)$  is parameterized as  in \eqref{eq:cov_k_t}; we denote  this model as \textsc{C-GWMF}. Lastly, we consider our proposed model, which adopts the flexible parameterization of covariance parameters $\rho(\cdot)$ and $\sigma(\cdot)$ given in \eqref{eq:basis_rho_mvar}. A summary of the models can be found in Table \ref{tab:models_summary},  and additional details on their specifications are given below.

\begin{table}[htb]
	\centering
	\caption{Model summaries and the parameters that relates to the different specifications of the latent field $u(\cdot)$.}
	\label{tab:models_summary}
	\begin{tabular}{p{2.4cm} p{8cm} p{3.5cm}}
		Model name & Description & $\vect\theta$ \\
		\midrule
		\textsc{WMF} & ``Stationary'' model that considers constant covariance parameters & $[\log\sigma_S\:\log\rho_S]\T$ \\
		\textsc{C-GWMF} & Covariance structure is described through a known covariate & $[b_1^\kappa\: b_2^\kappa\: b_1^\tau\: b_2^\tau]\T$ \\
		\textsc{B-GWMF} & Covariance structure is described as a sum of  $m_{\tilde{h}}$ basis functions & $[b^\sigma_1 \ldots b^\sigma_{\ncoarse} \: b^\rho_1 \ldots b^\rho_{\ncoarse} ]^\mathrm{T}$
	\end{tabular}
\end{table}

The \textsc{WMF} model defines
$u(\cdot)$ as the solution of \eqref{eq:spde_general} with constant covariance parameters: $\tau(\vect s)=\tau_S$ and $\kappa(\vect s)=\kappa_S$ in all $\vect s\in\Gamma$. Consequently, $\vect \theta=[\log\sigma_S\:\log\rho_S]\T$ contains only two parameters. This model is parsimonious and allows efficient estimation of the covariance parameters $\kappa_S$ and $\tau_S$, ( or even $\rho_S$ and $\sigma_S$ changing parameterization here using \eqref{eq:param_transformation}).
The covariance structure remains graph-dependent, and physical interpretation of the range and variance  parameters of the SPDE is only approximate. In particular, the covariance is Matérn-like in regions sufficiently far from vertices, while vertex conditions influence the covariance structure elsewhere.

For the \textsc{C-GWMF}, defined in \ref{eq:cov_k_t}, we assume that the covariance parameters depend on one single spatial covariate  $z(\vect s)$ known in all $\vect s\in\Gamma$.
In practice, however, a covariate that adequately explains the spatial covariance structure may be unavailable or difficult to identify. To reflect this setting, we assume that no such covariate is available and instead construct a surrogate covariate directly from the observed data.
Specifically,  we  use $y^r(\vect s^r_i)$ for all locations $s^r_i\in\mathcal S^r$ and $r=1,\ldots,\lfloor R/2 \rfloor$, where $\lfloor x \rfloor$ means rounded down to the closest integer, to construct a covariate, and the other half $y^r(\vect s^r_i)$ for locations $s^r_i\in\mathcal S^r$ and $r=\lfloor R/2\rfloor +1,\ldots, R$. 
Then $\vect\theta=[b^\kappa_1\:b^\kappa_2\:b^\tau_1\:b^\tau_2]\T$, so the number of hyperparameters going into the covariance structure is four when we include one covariate for both $\log\kappa(\cdot)$ and $\log\tau(\cdot)$.
This method follows \citet{Bolin2025nonstat}. Since the covariate is derived from data, it can capture relevant spatial spatial variation in the covariance structure. 
If the model assumptions are correct, one can construct a covariate closely related to the underlying marginal variance. This approach relies on a sufficient number of replicates $R$ as part of the data is used to build the covariate and the rest for the model fit. We will further look into how much data is needed for this approach in the rest of the simulation study. 

Finally, we consider the fully flexible GWM-GRF model, denoted \textsc{B-GWMF}, where the covariance parameters are represented using piecewise linear basis functions on a coarse mesh as in \eqref{eq:basis_rho_mvar}. This is the most flexible model as  as both $\rho(\cdot)$ and $\sigma(\cdot)$ can vary freely over the graph subject to the imposed regularization. Moreover, all data can be used to fit the model. Such flexibility comes at an increased computational cost. In addition to the fine mesh used for the latent field representation, a separate covariance mesh must be selected.

This additional mesh should be fine enough to capture relevant spatial variation in the structure,  but increasing resolution is linked to larger parameter vector  $\vect \theta=[b^\sigma_1 \ldots b^\sigma_{\ncoarse} \: b^\rho_1 \ldots b^\rho_{\ncoarse}]\T$ with dimension $2\ncoarse$. For computational reasons, this parameter vector should be kept as small as possible. The spatial design
is also a limitation to how fine the mesh can be, as the model cannot find any changes in the covariance structure without observations to support it.
The main computational challenge of this model is therefore estimating the large number of covariance parameters associated with the  coefficient $b^\rho_j$ for $j=1,\ldots,\ncoarse$ and $b^\sigma_j$ for $j=1,\ldots,\ncoarse$.

In the following simulation study, we consider the road network surrounding King Abdullah University of Science and Technology (KAUST), shown in Figure~\ref{fig:sstudy_meshes}. The network geometry was retrieved using \texttt{osmdata} \citep{cran-osmdata}, which provides access to OpenStreetMap data, and was subsequently processed in QGIS. The resulting graph comprises 159 vertices and 263 edges. Its diameter, defined as the maximal shortest-path distance between any pair of vertices, is \SI{4.8}{\kilo\meter}. The graph is for demonstration of a real-world road system represented as a metric graph, that has a natural boundaries where the network is dense in a certain area, with many intersections close in space, and is connected to the rest of the network through longer edges. The graph was easily available from \texttt{osmdata} and used for demonstration in the \texttt{MetricGraph} vignette.

\subsection{Scenarios}
\begin{figure}[htb]
	\centering
	\subfigure[Mesh with $h=0.5$\si{\kilo\meter}]{\includegraphics[width=0.45\linewidth]{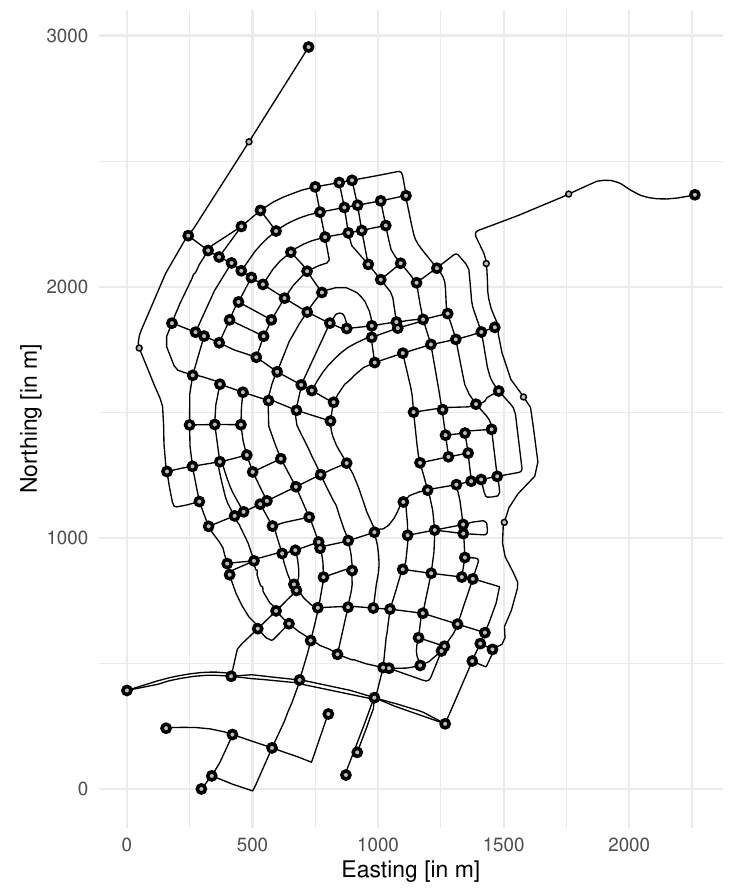}}
	\subfigure[Mesh with $h=0.1$\si{\kilo\meter}]{\includegraphics[width=0.45\linewidth]{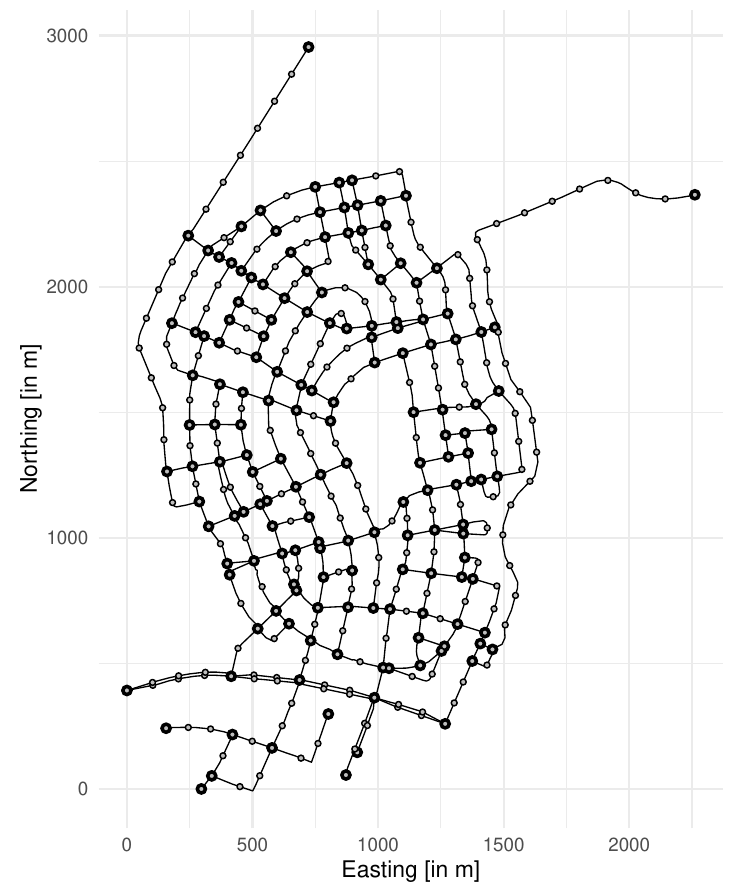}}
	\caption{The two meshes that were constructed to construct the three models, \textsc{WMF}, \textsc{C-GWMF} and \textsc{B-GWMF}. Note that the coarse mesh in (a) is only used in \textsc{B-GWMF}. Graph vertices, $v\in\mathcal V$ are black circles, and mesh vertices ($\mathcal V_h$ or $\mathcal V_{\tilde h}$) are smaller gray circles.}
	\label{fig:sstudy_meshes}
\end{figure}

Two computational meshes are constructed on the network:
(i) a coarse mesh with $\ncoarse=165$ vertices and a maximum spacing of $\tilde h=$ 0.5 \si{\kilo\meter}, and
(ii) a fine mesh with $\nfine=338$ vertices and a maximum spacing of $h=$ 0.1 \si{\kilo\meter}.
The coarse mesh contains six additional mesh vertices introduced along the longer edges, and locations are shown in Figure~\ref{fig:sstudy_meshes}. Shorter edges do not require additional vertices as they are already below the defined maximum spacing. Similarly, the fine mesh contains 179 additional vertices introduced according to the same refinement procedure.

\begin{figure}[htb]
	\centering
	\subfigure[$\eta^r(\vect s)$]{\includegraphics[width=0.45\linewidth]{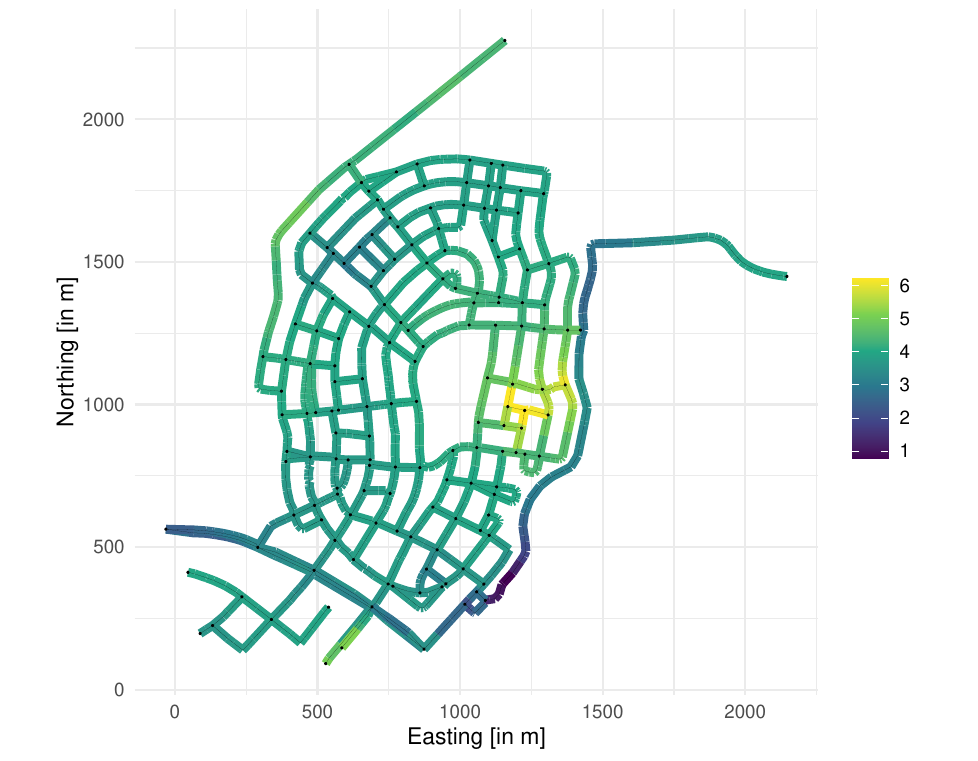}}
	\subfigure[$\vect y^r$]{\includegraphics[width=0.45\linewidth]{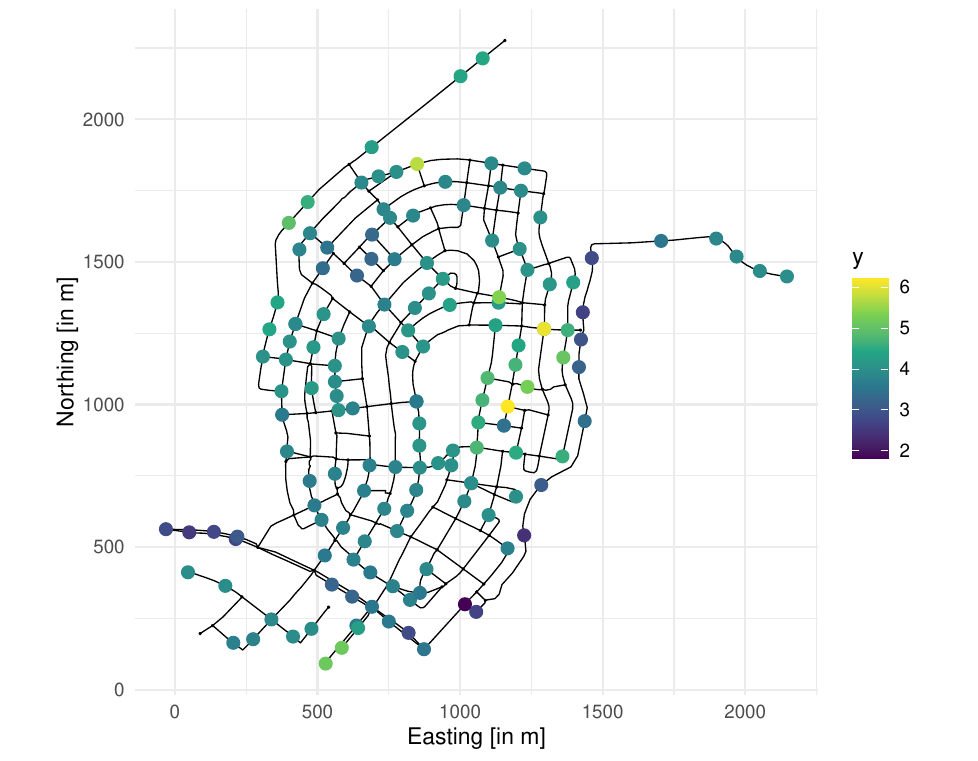}}
	\caption{A realization of (a) one replicate of the field, $\eta^r(\cdot)$, with true covariance structure as shown in Figure ~\ref{fig:sstudy_true_cov_fields} and (b) the set of observations, $\vect y^r$, related to that latent field with $N=150$.}
	\label{fig:sstudy_field_obs}
\end{figure}

We consider $R$ independent replicates of the same GRF on the graph $\Gamma$,
\begin{equation*}
	\eta^r(\vect s) = 4 + u^r(\vect s),\quad  \vect s\in\Gamma,r=1,\ldots, R,
\end{equation*}
where $4$ is the mean level  and $u^r(\cdot)$ are independent samples from the same distribution induced by \eqref{eq:spde_general} under the chosen parameterization of $\rho(\cdot)$ and $\sigma(\cdot)$.
For simplicity, we use the same observation locations for all replicates so that $\mathcal S^r=\mathcal S$, $\forall r$. As a consequence $N_r=N$, and the total number of observations is given by $n=N\cdot R$. 
Observation locations are sampled uniformly over the graph. Observations are defined by evaluating the replicates of $\eta(\cdot)$ in these locations and adding observation noise,
\begin{align*}
	y^r(\vect s_i) & \sim \mathcal{N}(\eta^r(\vect s_i),\sigma_\mathrm{N}^2),\quad i=1,\ldots, N, r=1,\ldots, R,
\end{align*}
where the observation noise variance is set to $\sigma_\mathrm{N}^2=0.01^2$. Figure~\ref{fig:sstudy_field_obs} shows one realization of the $\eta(\cdot)$ field in the non-stationary case and one example of the observation locations.

\begin{figure}[htb]
	\centering
	\subfigure[$\log\rho(\cdot)$]{\includegraphics[width=0.45\linewidth]{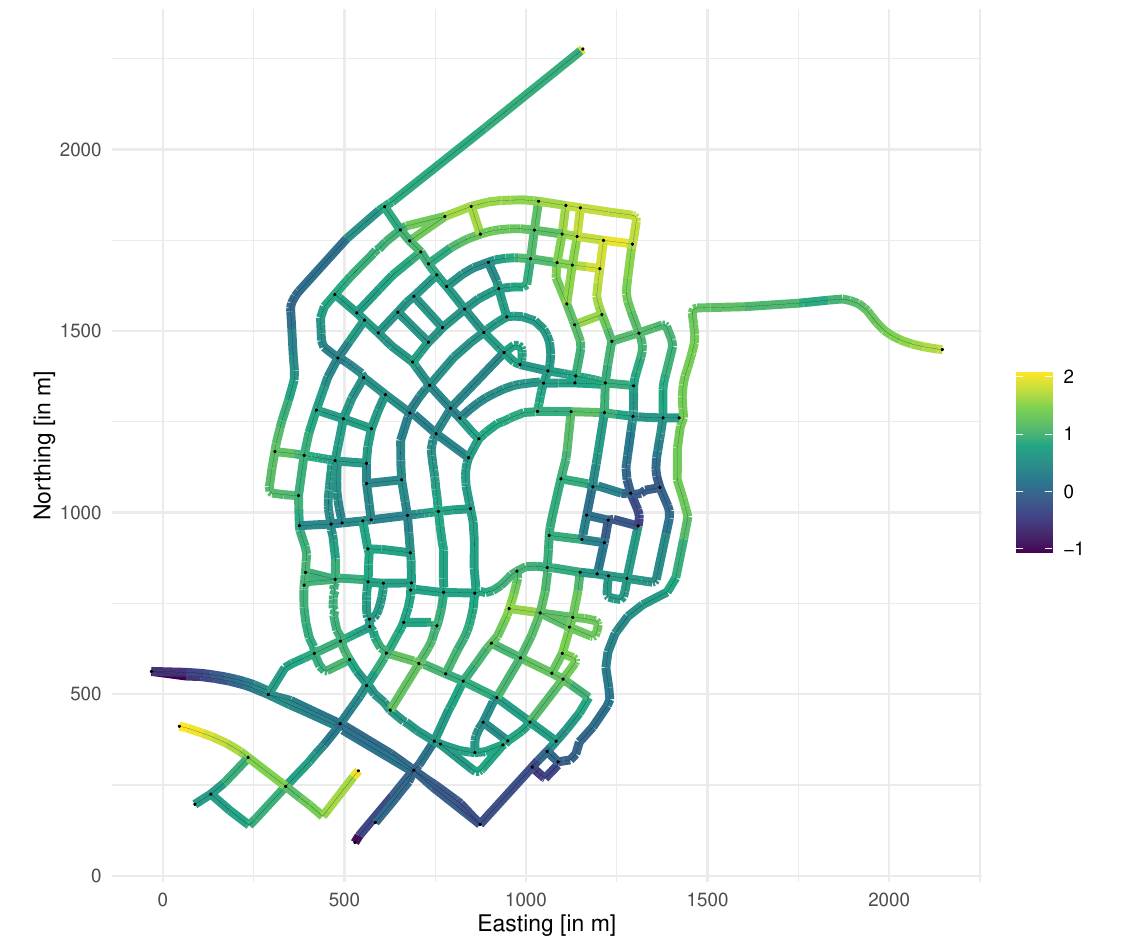}}
	\subfigure[$\log\sigma(\cdot)$]{\includegraphics[width=0.45\linewidth]{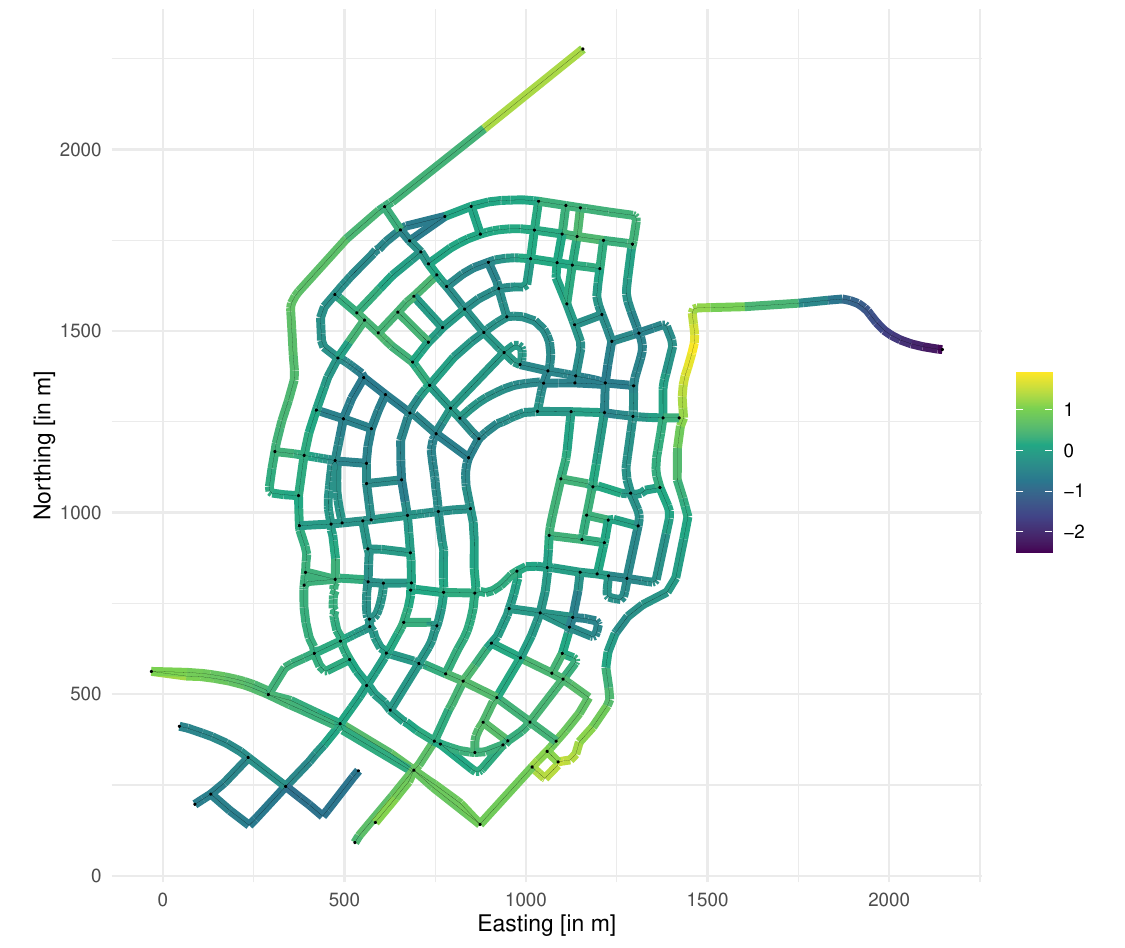}}
	\caption{True functions used for covariance structure parameterized with $\log\rho(\cdot)$ and $\log\sigma(\cdot)$.}
	\label{fig:sstudy_true_cov_fields}
\end{figure}

We consider two different scenarios for $u^r(\cdot)$. In the first scenario, $u^r(\cdot)$ is a stationary field with constant covariance parameters, $\log\rho(\vect s) = 0.6$ and $\log\sigma(\vect s) = 0.1$ for all $s \in \Gamma$. In the second scenario, $u^r(\cdot)$ is a non-stationary field whose range $\rho(\cdot)$ and standard deviation $\sigma(\cdot)$ vary over space. In this case, we simulate realizations of $\log\rho(\cdot)$ and $\log\sigma(\cdot)$ as independent WM-GRFs with approximate range 0.91 \si{\kilo\meter} and approximate standard deviation 0.92.
Thus, using the interpretation from the one-dimensional SPDE approach, point-wise deviations from the mean are typically of order 0.92 on the log-scale and at distance 0.91 \si{\kilo\meter} the correlation is near 0.1.
These parameters were chosen so that the covariance fields vary over the graph such that assumptions of \textsc{WMF} are too strict.
The coarse mesh is used to sample the true covariance structure.
Additionally, we add a constant to $\log\rho(\cdot)$ equal to $0.7$.
We keep the resulting realizations fixed and the realizations of $\log\rho(\cdot)$ and $\log\sigma(\cdot)$ are shown in Figure~\ref{fig:sstudy_true_cov_fields}.

We consider three values of $N$, corresponding to sparse ($N = 50$), medium ($N = 100$), and dense ($N = 150$) spatial coverage. The levels were chosen to represent cases where one is close to having as many observations as we have vertices, and down to quite few observations compared to the number of vertices and edges. To illustrate three levels of temporal coverage, we consider three values of $R$, namely 5, 25, and 50. The levels are chosen with our application to modeling of traffic data in mind. If one consider a specific time window with traffic data, for example Monday 10:00-11:00am, each of the resolution levels correspond to approximately one month, six months and a year.
For each of the 9 combinations of the resolution levels -- simulation setting -- we simulate 20 datasets that comes from randomly drawing observation locations from the simulated field, where each repeated experiment under the same simulation setting is independent of the next. For each dataset, we estimate the model parameters and predict the field $\eta(\vect s)$ in all vertices in $\mathcal V_h$. 

\subsection{Candidate models and evaluation}
For each of the two scenarios, we investigate how three models ---namely, \textsc{WMF}, \textsc{C-GWMF}, and \textsc{B-GWMF}---perform as a function of the number $N$ of spatial observation locations and the number $R$ of temporal replicates. We next describe the specification of these three models in more detail.

All three models share the same hierarchical structure described in Section~\ref{sec:model}. For the mean component we let $\vect\beta=[\beta_0]^\mathrm{T}$ and $\matr X=\vect 1_n$, where $\vect 1_n$ is a $n \times 1$ vector of ones. Thus, the mean structure only consist of an intercept.

For the \textsc{C-GWMF} model, we create the covariate needed to model the covariance structure from the data as follows: 
At each observation location $\vect s_i$, $i=1,\ldots,N$, we compute the empirical standard deviations of the observations $y^r(\vect s_i)$, $r=1,\ldots,\lfloor R/2 \rfloor$, by
\begin{equation*}
	z_d(\vect s_i) = \sqrt{\frac{1}{\lfloor R/2 \rfloor -1}\sum_{r=1}^{\lfloor R/2 \rfloor}(y^r(\vect s_i)- \bar{y}(\vect s_i))^2},\quad i=1,\ldots,N,
\end{equation*}
where $\bar{y}(\vect s_i)$ is the temporal average observed at station $\vect s_i$, across replicates $r=1,\ldots,\lfloor R/2 \rfloor$. Next, we create a continuous covariate by smoothing the log-transform of the empirical standard deviations, $\tilde{z}_d(\vect s_i)=\log(z_d(\vect s_i))$, $i=1,\ldots,N$. We smooth using a model where  $\vect{\tilde{z}}_d=[\tilde{z}_d(\vect s_1\:\ldots\: \tilde{z}_d(\vect s_N)]\T$ are noisy observations of a WM-GRF, $\tilde{z}(\cdot)$ as follows
$$
\tilde{z}_d(\vect s_i) = \tilde{z}_0 + \tilde{z}(\vect s_i) + \varepsilon_i\quad i=1,\ldots,N, \varepsilon_i\sim \mathcal N(0,\sigma_{N,z}^2),
$$
and $\tilde{z}(\cdot)$ can be described as the solution to the SPDE
$$
(\kappa_z^2-\Delta)(\tau_z \tilde{z}(\vect s)) =\mathcal W(\vect s) \quad \vect s\in\Gamma.
$$
We use the same mesh $\mathcal V_h$, and determine the precision matrix for weights $\vect {\tilde{z}}=[\tilde{z}(v_1)\:\ldots\: \tilde{z}(v_{\nfine})]\T$ and perform inference for the hierarchical model as described in the previous setting. 
The posterior mean is used as the covariate $\widehat{\tilde{z}}(\cdot)$ for both $\log\kappa(\cdot)$ and $\log\tau(\cdot)$ as was described in \eqref{eq:cov_k_t}. 
Note that, when fitting this model, we use only the second half of the observations $y^r(\vect s_i)$ for $i=1,\ldots,N$ and $r=\lfloor R/2\rfloor+1,\ldots, R$, as the first half of the data cannot go twice into the model. That is, we get $\vect Y =[{\vect (y^{\lfloor R/2\rfloor+1}})\T\:\ldots\: (\vect {y}^R)\T]\T$, and $n=N\cdot (R-\lfloor R/2\rfloor)$. 
The hierarchical model takes the form as we saw in \eqref{eq:hierarchical_model} with $\matr Q_{h,q}=\matr Q_{h,2}$ and we only consider $\vect w^r$, $r=\lfloor R/2\rfloor+1,\ldots, R$.


Choosing reasonable priors directly for $\kappa(\cdot)$ and $\tau(\cdot)$ is challenging, since these parameters do not have a direct physical interpretation. Instead, we recommend specifying prior information in terms of the approximate correlation range and marginal standard deviation using the transformations in \eqref{eq:param_transformation}. Prior means for $\rho(\cdot)$ and $\sigma(\cdot)$ can then be transformed to the corresponding values of $\kappa(\cdot)$ and $\tau(\cdot)$ on the log scale. In this parameterization, the regression coefficients associated with the covariate, $b_2^\kappa$ and $b_2^\tau$, naturally have prior mean zero, corresponding to no covariate effect.
In the simulation study, the same prior is used for all simulated datasets and in both scenarios. The prior is specified using knowledge of the true covariance structure employed to generate the data and is not adapted to the individual datasets used for model fitting.


For both the \textsc{WMF} and \textsc{B-GWMF} model, the whole dataset is used for inference on model parameter. Therefore, $\vect Y=[{\vect y^1}\T\:\ldots\:{\vect y^R}\T]\T$ and $n=N \cdot R$, and the hierarchical model is as stated in \eqref{eq:hierarchical_model}.

The prior for $\vect \theta$ in the simplest model \textsc{WMF} uses a domain-based prior for $\rho_S$.
We set the median as 0.3 times the diameter of the bounding box of the graph.
This gives a median of about 1\si{\kilo\meter} for $\rho_S$ (be aware that the priors are set for $\log\rho_S$). For $\sigma_S$ we choose a prior median of 1 (0 on log-scale). The precision is 0.1 for both $\log\kappa$ and $\log\tau$. 

For \textsc{B-GWMF}, we use multivariate Gaussians with precision matrix as described in \eqref{eq:mod_Q} with $\tau_1=0.01$ and $\tau_2=1$. 
We follow our suggested approach and consider different values of $\tau_1$ while $\tau_2=1$ is kept fixed, sample 100 times from $N(0,\matr Q(\vect\theta)^{-1})$ and compute the 95\%-quantiles across these samples. By visualizing the field and comparing these to the true covariance structure in Figure \ref{fig:sstudy_true_cov_fields}, we set $\tau_1=0.01$. Note that we use the same prior in both scenarios, including the spatially constant truth. In addition, we have a fixed offset parameter, $\log\rho_0$ (similar to $f_0$ in \eqref{eq:general_f}), which we set as $0.5$, which is intentionally not equal to the true offset equal to 0.7. That is, we still have to rely on \textsc{B-GWMF} to correct the level of $\log\rho(\cdot)$ through the basis function weights $b^\rho_j$, $j=1,\ldots, \ncoarse$. 

We evaluate the models' ability to recover the log-transformed range $\log\rho(\vect s)$ and log-transformed standard deviation $\log\sigma(\vect s)$ of the Whittle--Mat\'ern fields $u^r(\vect s)$, $r=1,\ldots, R$ through root mean squared error (RMSE). Their ability to predict the field $\eta^r(\vect s)$, $r=1,\ldots, R$, at unobserved locations is evaluated with both RMSE and the continuous ranked probability score (CRPS). While the covariance parameter RMSE measures how well each model recovers the underlying covariance structure, the predictive scores quantify the quality of the latent field predictions regardless of whether the covariance parameters are recovered exactly. Since the data are simulated, both the true covariance fields and the latent fields are available, allowing all evaluation metrics to be computed directly. 
Differences between the estimated and true covariance structure at each mesh node of the coarser mesh $\mathcal V_{\tilde h}$ are assessed directly and RMSE is defined as:
\begin{equation*}
	\text{RMSE}_f=\sqrt{\frac 1 R\sum^R_{r=1} \frac{1}{\ncoarse} \sum_{\vect s_k\in\mathcal V_{\tilde h}} (f(\vect s_k) - \hat f(\vect s_k))^2},
\end{equation*}
where $f(\cdot)$ is either $\log\rho(\cdot)$ or $\log\sigma(\cdot)$. For \textsc{WMF} $\hat f(\cdot)$ is just a constant, while for \textsc{C-GWMF} we compute the transformations using \eqref{eq:param_transformation}. \textsc{B-GWMF} estimates the field directly on this scale.

We predict the latent field at each node of the finer mesh $\mathcal V_h$ and define the RMSE for the latent field as:
\begin{equation*}
	\text{RMSE}_\eta=\sqrt{\frac 1 R\sum^R_{r=1} \frac{1}{\nfine} \sum_{\vect s_k\in\mathcal V_h} (\eta^r(\vect s_k) - \hat\eta^r(\vect s_k))^2}.
\end{equation*}
where $\hat\eta^r(\vect s_k)$ is the posterior mean of $\eta^r(\vect s_k)$.
Similarly for CRPS, we compute
\begin{equation*}
	\text{CRPS}_\eta=\frac 1 R\sum^R_{r=1} \frac{1}{\nfine} \sum_{\vect s_k\in\mathcal V_{h}} \text{crps}(\eta^r(\vect s_k), \hat\eta^r(\vect s_k), \text{sd}(\hat\eta^r(\vect s_k))),
\end{equation*}
where $\text{sd}(\hat\eta^r(\vect s_k))$ is the posterior standard deviation associated with the prediction $\hat\eta^r(\vect s_k)$, and
\begin{equation}\label{eq:crps_normal}
	\text{crps}(y, \mu, \sigma) = \sigma\left[\frac{y-\mu}{\sigma}\left[2\Phi\left(\frac{y-\mu}{\sigma}\right)-1\right] + 2\phi\left(\frac{y-\mu}{\sigma}\right) -\frac{1}{\sqrt \pi}\right],
\end{equation}
where $y$ is the observation, $\mu$ is the predicted value from the model and $\sigma$ is the model uncertainty.

\subsection{Results}

We first present results for the WM-GRF scenario, where the true field has constant covariance parameters. In this case, the three models exhibit similar prediction performances for the latent field $\eta^r(\vect s)$, as measured by RMSE and CRPS when the number of replicates $R$ is equal or above 25, which is shown in Figure~\ref{fig:rmse_crps_stat}(a) and Figure~\ref{fig:rmse_crps_stat}(b). 
For the smallest number of replicates, however, \textsc{B-GWMF} performs somewhat worse than the two competing models. We note that \textsc{B-GWMF} uses the same prior specification in both simulation scenario considered. In practice, one could/should set a stricter prior when prior knowledge or exploratory analysis suggests that the covariance structure is (close to) spatially constant. Figures can be found in Appendix~\ref{app:B}. 

When it comes to comparing the ability to recover the true covariance structure, \textsc{WMF} achieves the lowest RMSE across all simulation settings, with \textsc{C-GWMF} performing similarly, as can be seen in Figure~\ref{fig:rmse_range_sdev_stat}(a) and Figure~\ref{fig:rmse_range_sdev_stat}(b). 
For $R=50$, the models are consistently ranked, from  best to worst  with regards to RMSE for the covariance fields as \textsc{WMF}, \textsc{C-GWMF} and \textsc{B-GWMF}. 
This reflects the bias--variance trade-off: when the true covariance structure is spatially constant, the additional flexibility of \textsc{C-GWMF} and \textsc{B-GWMF} is unnecessary and can lead to the estimation of spurious spatial variation in the covariance parameters. Consequently, the simpler \textsc{WMF} model provides the most accurate recovery of the true covariance structure. 
Detailed results are provided in Appendix~\ref{app:B}.

\begin{figure}[htb]
	\centering
	\subfigure[RMSE\label{fig:simstudy_rmse}]{\includegraphics[width=0.45\linewidth]{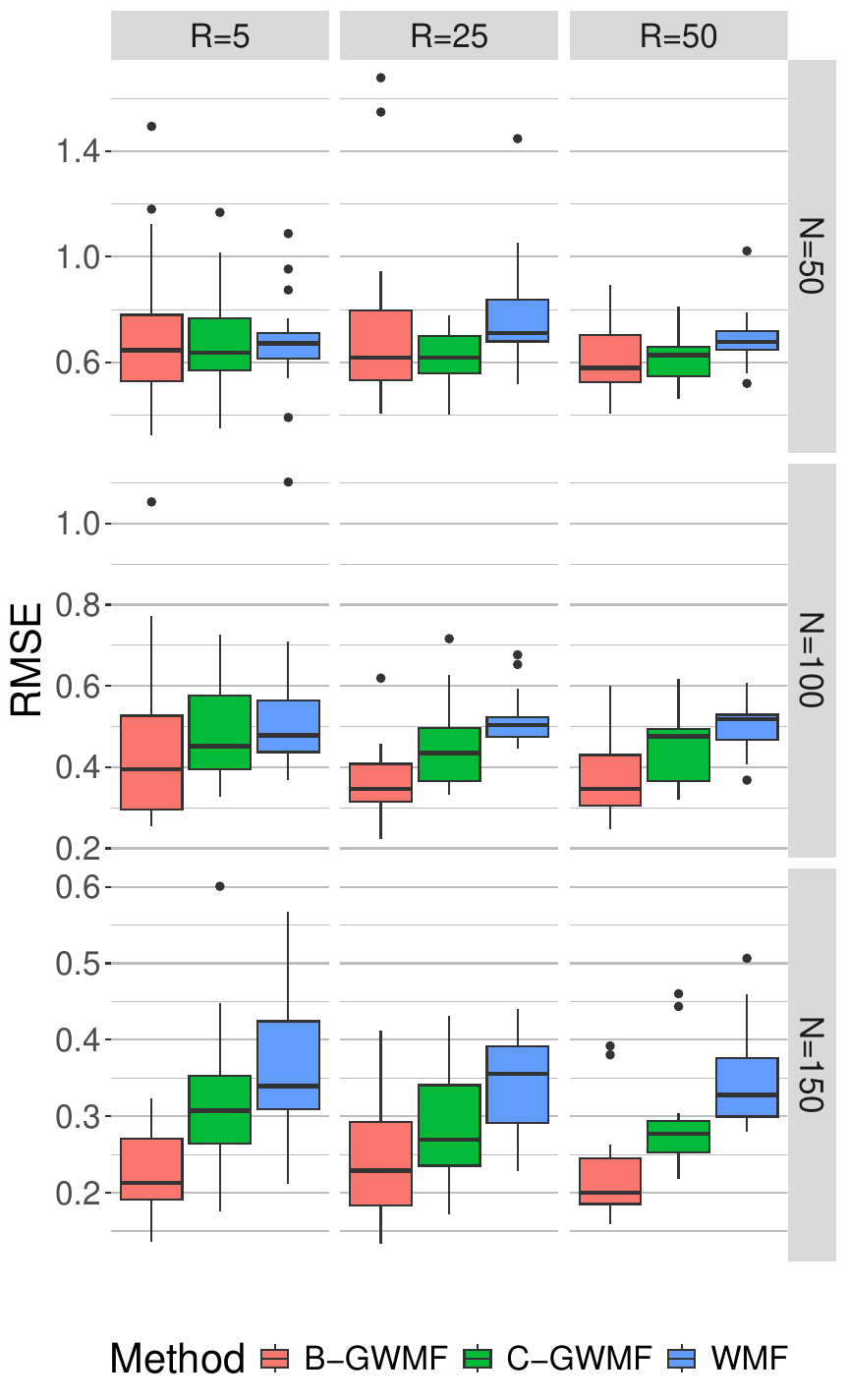}}
	\subfigure[CRPS\label{fig:simstudy_crps}]{\includegraphics[width=0.45\linewidth]{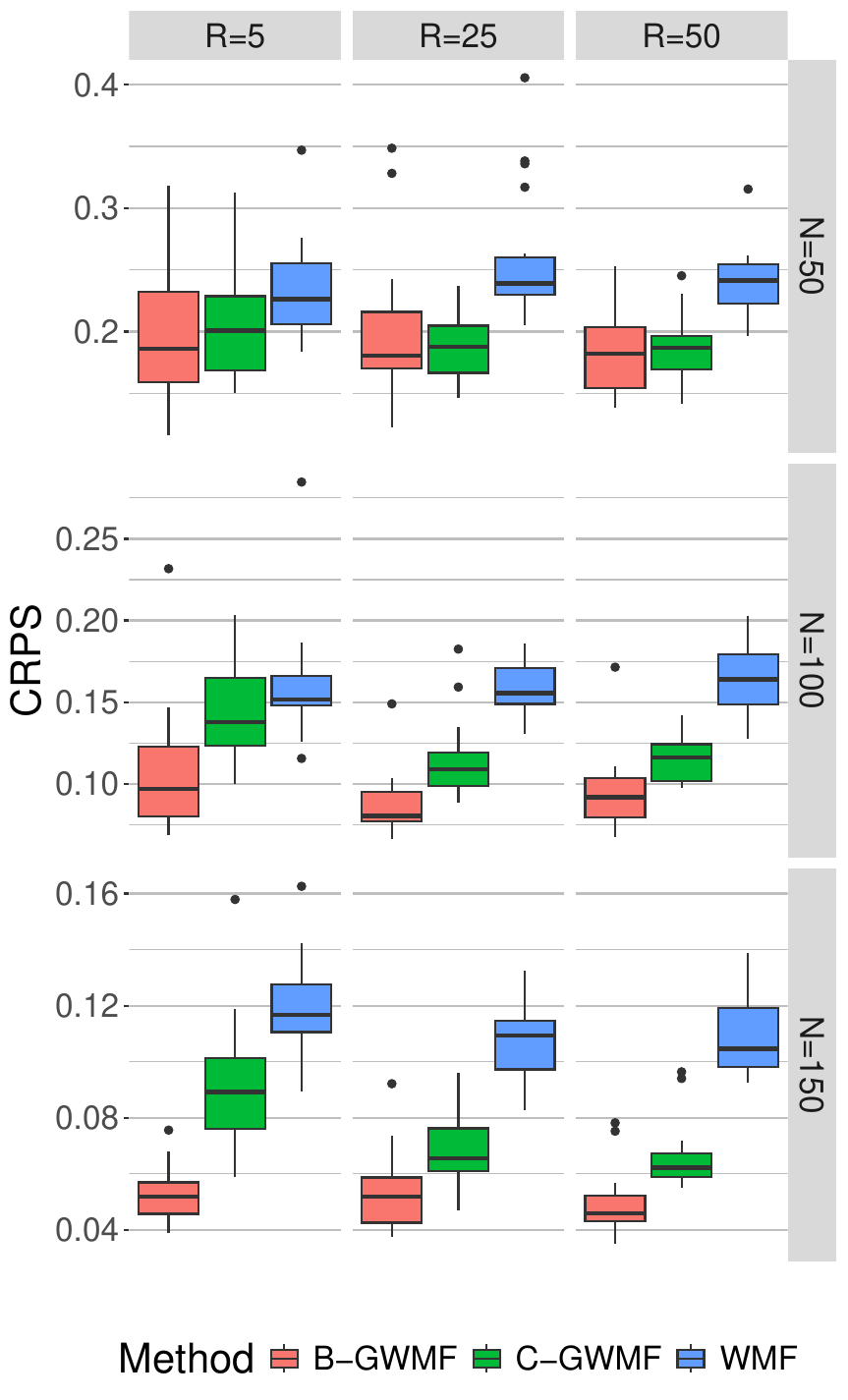}}
	\caption{RMSE (a) and CRPS (b) computed in all mesh vertices in the fine mesh for the latent field for all levels of spatial coverage (column) and resolution of repeated measurements/replicates (row) in the scenario with true covariance structure depicted in Figure~\ref{fig:sstudy_true_cov_fields}. The y-axis is scaled to better see the result in all simulation settings; two points are removed for \textsc{B-GWMF} in the setting with lowest spatial coverage ($N=50$) and temporal resolution $R=5$ (RMSE scores 4.26 and 4.56 and CRPS 1.79 and 0.75, respectively), and two points from \textsc{B-GWMF} with $N=50$ and $R=50$ (RMSE scores 4.25 and 16.5 and CRPS 0.539 and 1.49, respectively).}
	\label{fig:simstudy_rmse_and_crps}
\end{figure}

We next consider results for the more challenging GWM-GRF scenario with covariance parameters  vary in space as illustrated in   Figure~\ref{fig:sstudy_true_cov_fields}. We first consider the prediction of the field $\eta^r(\vect s)$. As expected,  increasing the number of spatial observation locations $N$ reduces both the RMSE and CRPS for the two models that can estimate variation in the covariance structure (row-wise in Figure~\ref{fig:simstudy_rmse} and Figure~\ref{fig:simstudy_crps}). Similarly, performance improves as the number of replicates increases. Overall, for $N \ge 100$ \textsc{B-GWMF} consistently provides the best predictive performance.
When the number of both replicates ($R$) and spatial locations ($N$) is small, all models show similar performances, with a slight preference for \textsc{B-GWMF}. This is expected as, in this setting, the available data provide limited information about the spatially varying covariance structure, making it difficult for the more flexible models to exploit their additional degrees of freedom.
When either $N$ or $R$ increases, the benefit of modeling spatially varying covariance parameters becomes increasingly apparent. This improvement is particularly evident in the CRPS results, where \textsc{B-GWMF} clearly outperforms the competing models already for (N=100).

\begin{figure}[htb]
	\centering
	\subfigure[RMSE$_{\log\rho}$\label{fig:simstudy_rmse_range}]{\includegraphics[width=0.45\linewidth]{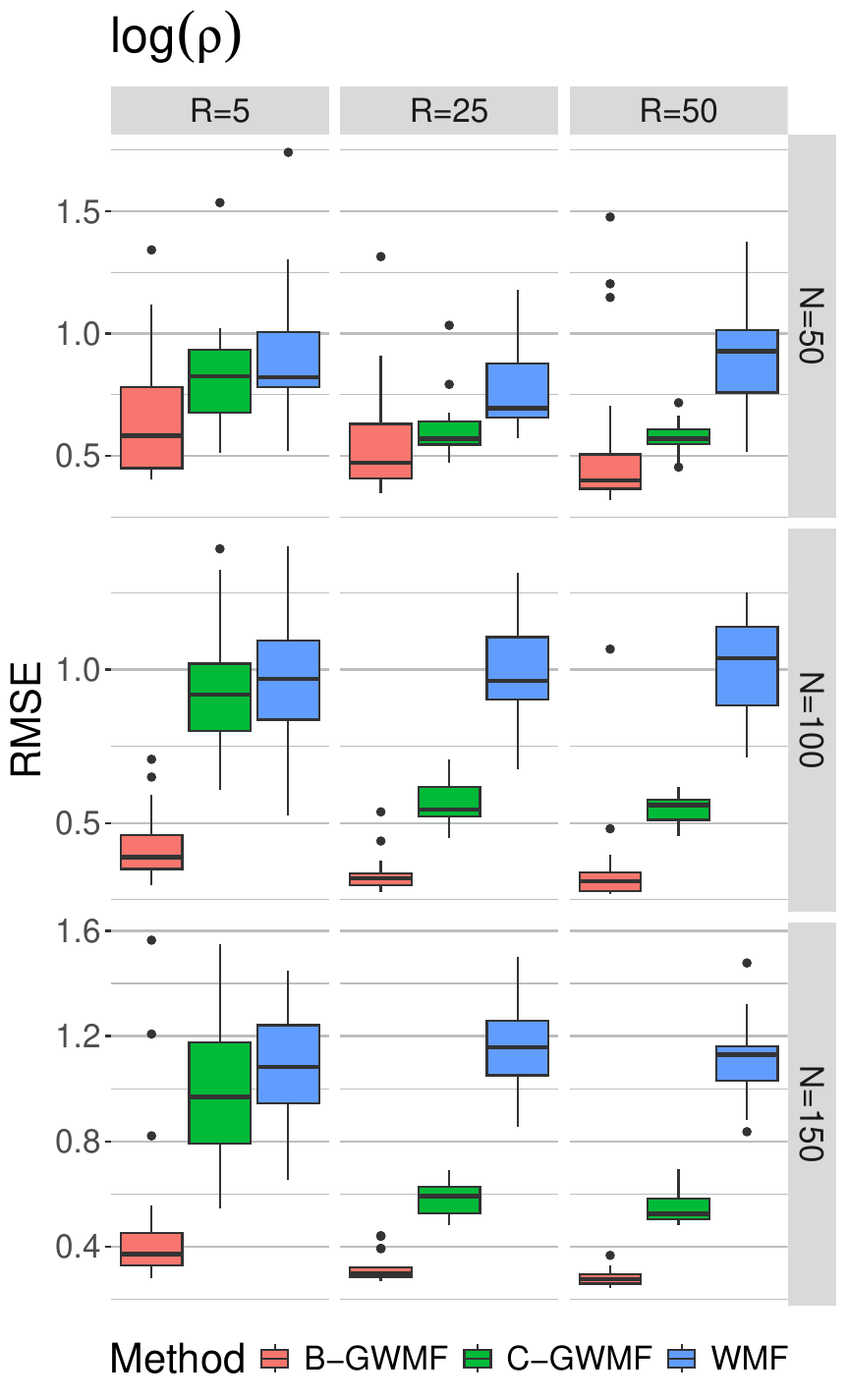}}
	\subfigure[RMSE$_{\log\sigma}$\label{fig:simstudy_rmse_stdev}]{\includegraphics[width=0.45\linewidth]{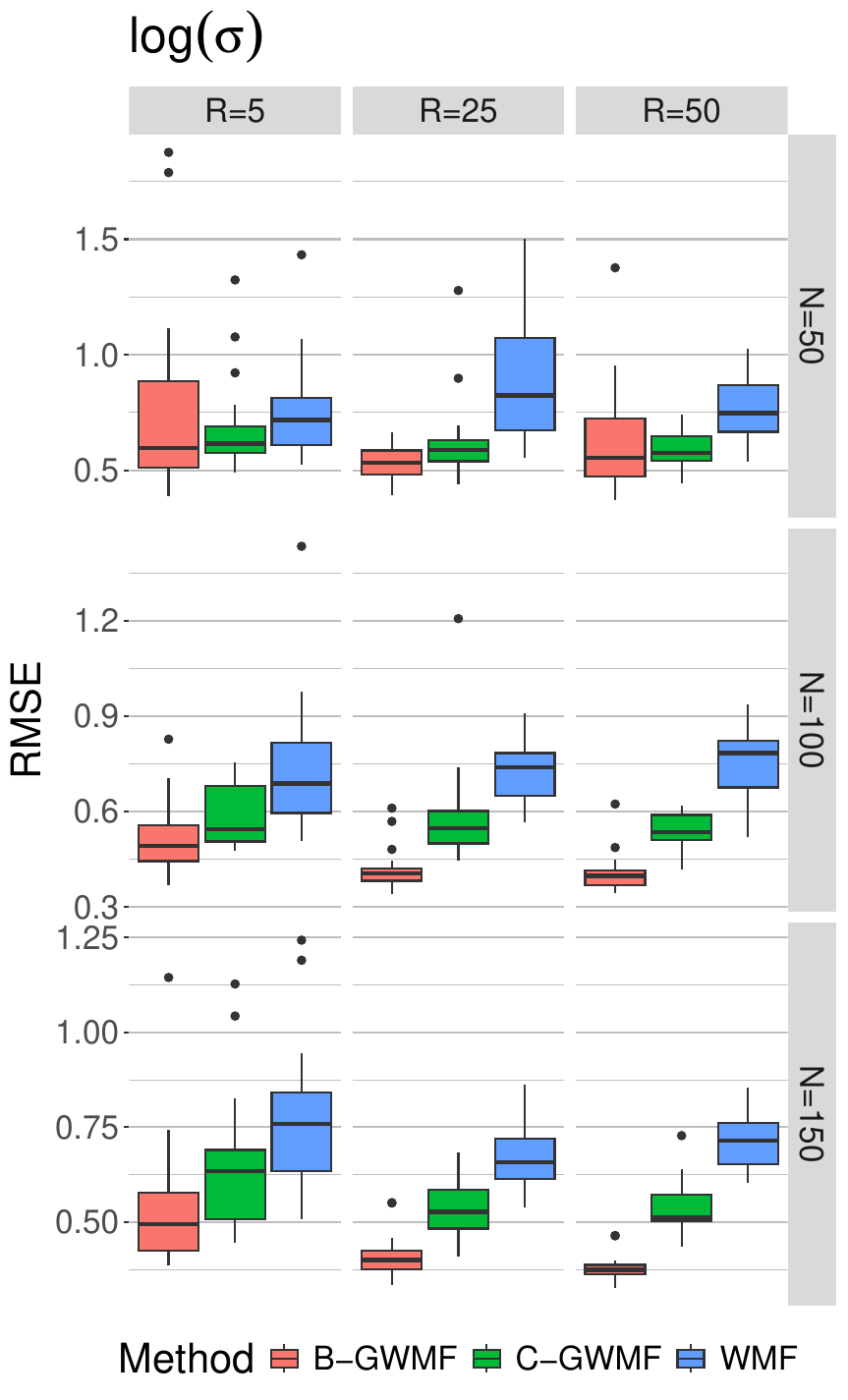}}
	\caption{RMSE computed in all mesh vertices in the coarse mesh for the log-transform of the covariance field of range for all levels of spatial coverage (column) and resolution of repeated measurements/replicates (row) in the scenario with true covariance structure as depicted in Figure~\ref{fig:sstudy_true_cov_fields}. For better visualization, two points in (a) were removed for \textsc{C-GWMF} in the setting with $R=5$ and $N=50$ (RMSE of 4.56) and $R=5$ and $N=100$ (RMSE of 3.60) and in (b) two points were removed for \textsc{C-GWMF} in the setting with $R=5$ and $N=50$ (RMSE of 6.89) and $R=5$ and $N=100$ (RMSE of 3.20) and one point was removed for \textsc{B-GWMF} in the setting $R=5$ and $N=150$ (RMSE of 2.16).}
	\label{fig:simstudy_rmse_range_and_stdev}
\end{figure}

The recovery of the true covariance structure is assessed using the RMSE of the estimated $\log\rho(\cdot)$ and $\log\sigma(\cdot)$ fields, shown in Figure~\ref{fig:simstudy_rmse_range} and Figure~\ref{fig:simstudy_rmse_stdev}. For the range field $\log\rho(\cdot)$,  \textsc{B-GWMF} consistently achieves the lowest RMSE,  and the difference becomes more clear under higher temporal ($R$) and spatial ($N$) coverage. In contrast, increasing the amount of data has little effect on \textsc{WMF} as the model will only estimate the field by one single parameter. This is reflected in the nearly constant RMSE across values of $R$ for the \textsc{WMF} model (Figure \ref{fig:simstudy_rmse_range_and_stdev}).
When $R\ge 25$, we see a clear improvement in recovering the range field for both \textsc{C-GWMF} and \textsc{B-GWMF}. The latter is outperforming the other models already with $N=100$ and $R=5$. 

For  the marginal standard deviation field, all models show similar performances when spatial- and temporal coverage is low. A clear difference between the models can be seen in settings with $R\ge25$ and $N\ge100$. See Figure \ref{fig:simstudy_rmse_range_and_stdev}. 

From our simulation study we conclude that, 
when the covariance structure is spatially constant, the complex models can achieve similar prediction performances to \textsc{WMF} provided that sufficient data are available  ($R\ge 25$). However, the identification of the true covariance structure is less successful, particularly for \textsc{B-GWMF}. This effect could probably be mitigated by having stricter priors ($\tau_1$ larger) than the one we considered here. To facilitate a fair comparison across all  simulation settings, we intentionally used the same prior specification for every model and scenario.

On the other side, when the covariance structure is spatially varying, the flexible models become increasingly advantageous as the amount of data increases. 
In our simulation study, $R=5$ replicates provide limited information for reliably estimating the covariance fields, whereas the estimates improve substantially for $R\geq25$. Under sufficiently dense spatial sampling ($N\geq100$), \textsc{B-GWMF} consistently provides the most accurate recovery of the covariance structure and the best predictive performance at unobserved locations. These results demonstrate that modeling spatially varying covariance parameters is beneficial when supported by adequate temporal and spatial information.


\section{Case study}\label{sec:case}
\subsection{Objective}
We consider traffic intensity measurements from fixed sensors in Madrid, Spain, observed during a recurring morning rush-hour window. The purpose of the case study is not to model the full temporal dynamics of traffic, but to study the spatial distribution of log traffic intensity (number of cars passing within a specified time window) on a road network under a fixed temporal regime. 

This provides a natural application for metric graph models: the observations are tied to a road network, and dependence is expected to follow network connectivity rather than Euclidean distance alone. Madrid is chosen because of the extensive availability of traffic data. The city has a large network of 4,962 fixed measuring stations with measurements available at 15-minute intervals and historical records extending back to 2013.
By restricting the analysis to a smaller district in Madrid, we obtain a graph that is of similar size to the graph we considered in the previous setting. 

We assume that for a narrow time window (1 hour), one can determine a true underlying spatial field for the log-transformed traffic intensity that describes the distribution of vehicles on the network. Traffic flow is influenced by the number of vehicles present on a road segment during a given time window. This field can give information on the spatial distribution of vehicles in the network.

We compare the three models studied in Section~\ref{sec:simstudy} on the chosen dataset. The models are evaluated based on in-sample prediction scores for the second half of the replicates, as \textsc{C-GWMF} only uses this partition of the data to fit the full model with covariance parameters that are spatially varying.
Leave one out cross-validation (LOOCV) is done to evaluate conditional predictions of unseen observations using the estimated covariance structure and held-out replicates. 
Regular cross-validation methods are infeasible with the flexible model on the graph considered in this case, as the computations of fitting the model are too time-consuming. 
We therefore use the estimated model parameters estimated from the training dataset, $\hat\beta$ and $\hat\theta$, and evaluate the models' predictive power on another dataset using the conditional distribution of $y^r_i \mid \vect y^r_{-i}, \vect\beta=\hat {\vect\beta},\vect\theta=\hat{\vect \theta}$, where $y^r_i$ refers to the left-out station and $\vect y^r_{-i}$ is the set of observations from the other stations of replicate $r=1,\ldots, R_{*}$, and $\hat\beta$ and $\hat{\vect\theta}$ are the model parameters that have been estimated with the training dataset.

\subsection{Data}
The traffic station data from Madrid, Spain, was obtained from \url{datos.madrid.es}, an open data portal managed by the City Council of Madrid. We consider months January and February from 2023 to 2026. We limit the dataset to the narrow time interval of Mondays 08:00-09:00am. In the original dataset, traffic counts are summed over 15-minute periods and multiplied by 4 to provide intensities in vehicles per hour (veh/h).
We summarize measurements within the same hour to obtain traffic counts for the full hour based on the 15-minute measurements. If the counts are zero for the whole hour, they are removed from the dataset because we assume that these low counts are a result of the sensor not working. Additionally, we remove observations from Mondays in the first week of the year, which includes national holiday days in Spain.

\begin{figure}
	\centering
	\subfigure[Map of Madrid, Spain with districts\label{fig:districts_of_madrid}]{\includegraphics[width=0.38\linewidth]{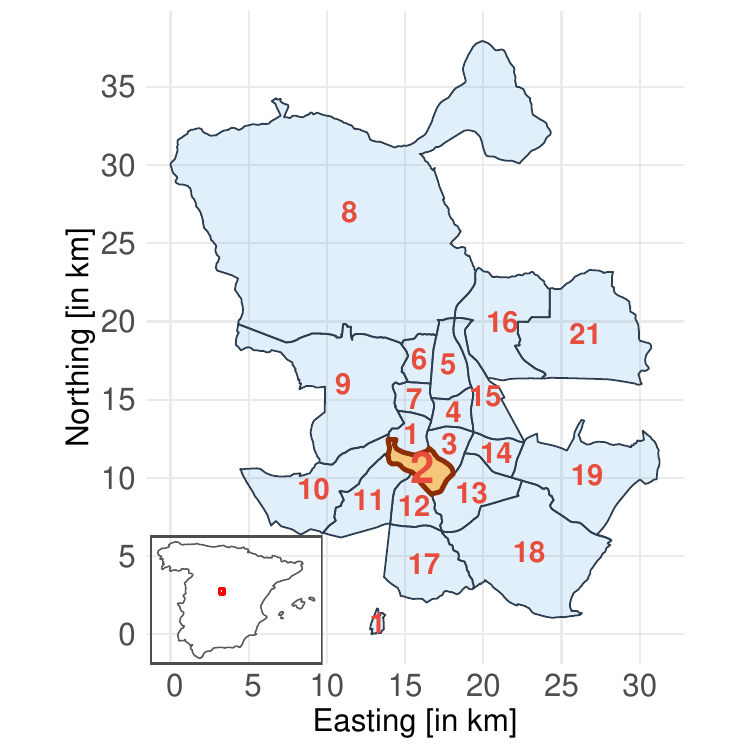}}\\
	\subfigure[Observed station mean\label{fig:madrid_observed_mean}]{
		\includegraphics[width=0.4\textwidth]{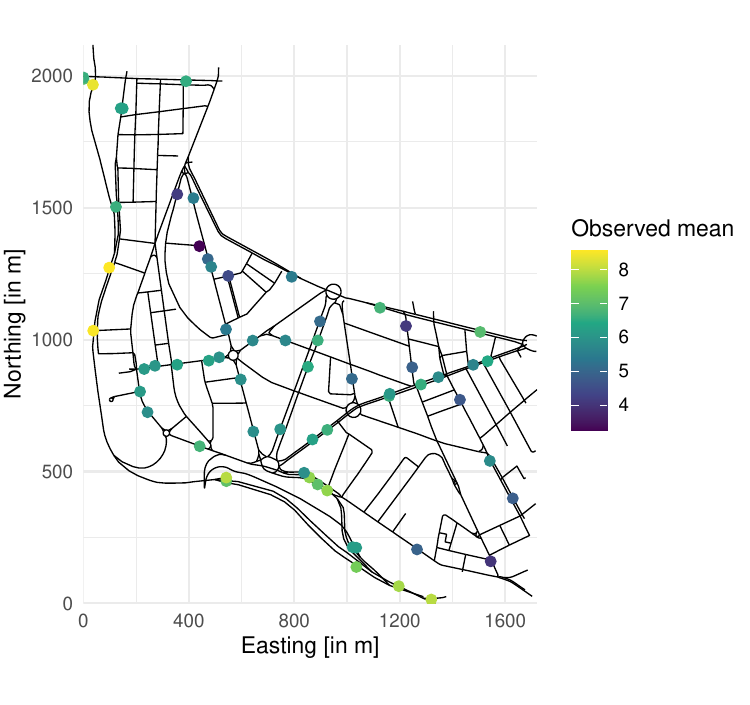}
	}
	\subfigure[Road type covariate\label{fig:covariate_for_mean}]{
		\includegraphics[width=0.4\textwidth]{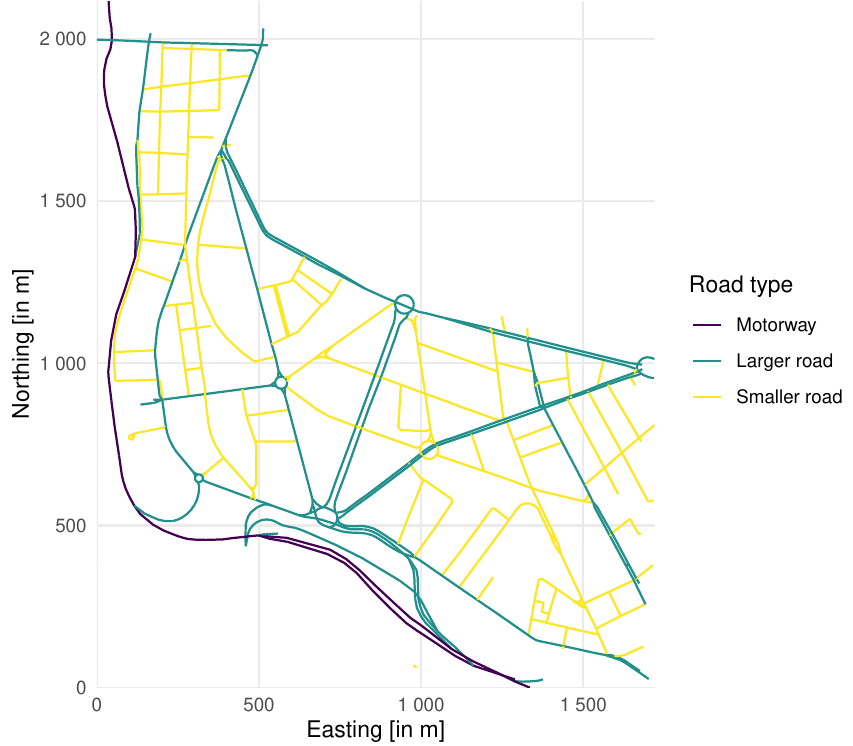}
	}
	\subfigure[Observed station st.dev.\label{fig:madrid_observed_sd}]{
		\includegraphics[width=0.4\textwidth]{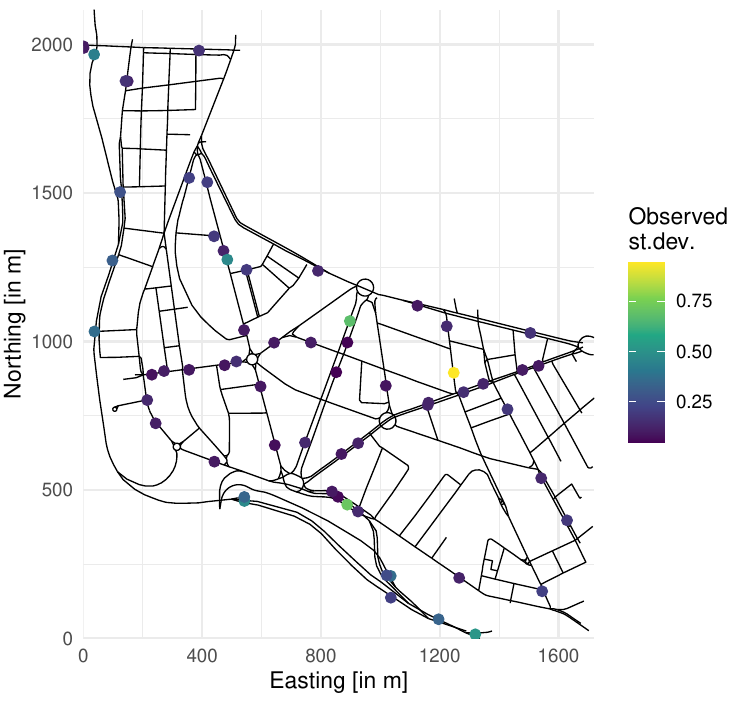}
	}
	\subfigure[Covariate for \textsc{C-GWMF}\label{fig:logsd_for_cgwmf}]{
		\includegraphics[width=0.4\textwidth]{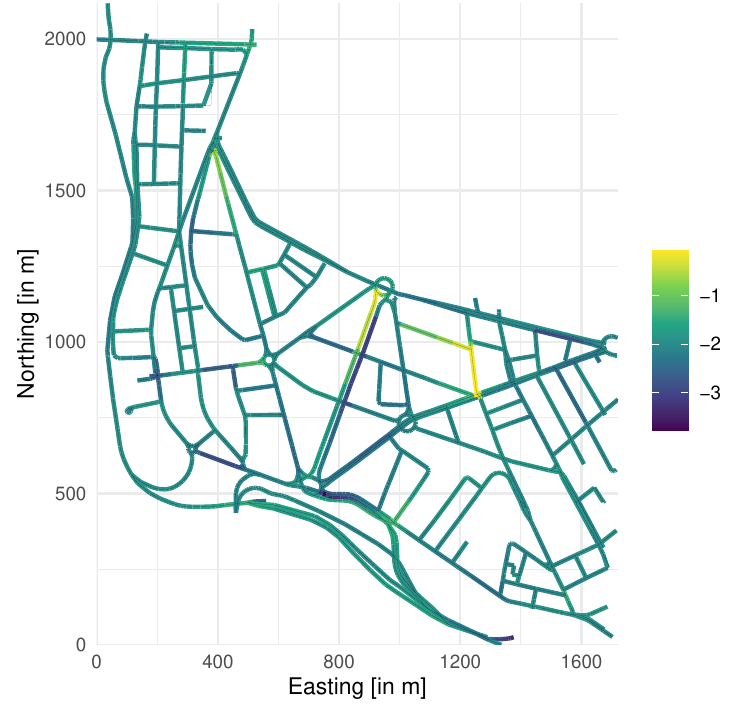}
	}
	\caption{A map of (a) districts of Madrid, Spain, together with the road network we consider with (b) average log-transformed traffic count data from traffic stations, (c) categorization of the different road types, (d) empirical standard deviation of log-transformed traffic count data and (e) the covariate constructed to be used in the covariance structure with \textsc{C-GWMF}.}
	\label{fig:districts_data_cov}
\end{figure}

The graph is obtained from \texttt{osmdata} \citep{cran-osmdata}.
The 21 districts of Madrid are shown in Figure~\ref{fig:districts_of_madrid}. 
To make the \textsc{B-GWMF} model computationally feasible we need to select a smaller area of the city.
First, we exclude larger districts like 8-21 because of the size of the domains. Further, after some exploratory data analysis where we consider the empirical standard deviation at the different sensors, 
we limit the study to district 2.
This graph proved to be too large for the  \textsc{B-GWMF}, therefore we further shrinked the study area.
The resulting graph, shown in Figure~\ref{fig:districts_data_cov}, has 270 vertices and 385 edges after pruning, with a diameter of 3.6 \si{\kilo\meter}. 
It is larger, in terms of number of vertices and edges, than the one considered in Section~\ref{sec:simstudy}. It is also ``cut'' more arbitrary than the graph we studied before, where to the East, it is simply cut from a specified bounding box (maximal longitude).

We split the dataset into a training set,  used to fit the three models and estimate covariance parameters, and a test set, used to evaluate the out-of-sample prediction abilities of the models. The training set covers the years 2023 to 2025, while the test set includes data from 2026. 

The training data comprises 62 unique traffic sensors $\mathcal S=\{\vect s_i\}_{i=1}^{62}=\bigcup_r \mathcal S^r$ and a total of 1,411 observations. Here, $\mathcal S^r\subset \mathcal S$, denotes the sensor locations observed in  replicate $r$ with  $\lvert \mathcal S^r\rvert=N_r$. Stations with missing counts or counts of zero or one are excluded. Then the training set contains  $N=\max_r (N_r)=62$ unique sensors and $R=23$ replicates. In terms of the data coverage settings  considered in Section~\ref{sec:simstudy}, this corresponds to low to medium spatial coverage and medium temporal resolution. Note that a larger graph also increases the computational complexity of the models.

Figure~\ref{fig:madrid_observed_mean} shows the average station log-intensity for 2023-2025. To account for systematic differences in mean intensity between roads, we use the road classification available for the spatial lines in 
\texttt{osmdata}. 
We collapse the original categories into three groups:
\textit{motorway}, \textit{larger road} and \textit{smaller road},  according to Table~\ref{tab:roadtypes}. The resulting covariate is shown in Figure~\ref{fig:covariate_for_mean}. For the graph considered here, the  ``trunk\_link'' and ``unclassified'' highway categories were not required to obtain a fully connected graph:this might be different for  for other road networks.

\begin{table}[htb]
	\centering
	\caption{We construct a covariate that identifies road type into three levels. This is used to allow for different means for each category.}
	\begin{tabular}{l >{\raggedright\arraybackslash}p{0.3\textwidth}}
		New category & Highway categories in \texttt{osmdata}  \\
		\midrule
		Motorway & ``motorway'' \\
		Larger road & ``motorway\_link'', ``primary'', ``secondary'', ``tertiary''  \\
		Smaller road &  ``residential'' \\
	\end{tabular}
	\label{tab:roadtypes}
\end{table}

In the test set, we use the same stations as in the training set, 
although the number of observed stations may again vary between replicates.
We denote the prediction locations in replicate $r$  by $\mathcal S^r_{*}=\{\vect s^r_k\}_{k=1}^{K_r}$ where $r=1,\ldots,R_{*}$ and  $R_{*}=7$ is the number of test-set replicates.

\subsection{Candidate models and evaluation}
We consider the same three models as in Section~\ref{sec:simstudy} and assume the following hierarchical structure,
\begin{align*}
	y^r(\vect s^r_i) &= \eta^r(\vect s^r_i) + \varepsilon_i^r,\quad \vect s^r_i \in \mathcal S^r, r=1,\ldots,R, \\
	\eta^r(\vect s) &= \vect x (\vect s)\T \vect \beta + u^r(\vect s), \quad \vect s\in\Gamma
\end{align*}
where $y^r(\vect s^r_i)$ is a noisy observations of the  underlying log-traffic intensity $\eta^r(\vect s_i)$, in location $\vect s^r_i\in\mathcal S^r$. 
The latent  field includes  linear covariate effects and a spatially varying field $u^r(\cdot)$. 
The fixed component and the same for all three models and 
we set
$\vect x(\vect s)=[1 \:\vect x_\text{rt}(\vect s)\T]\T$ and
$\vect x_\text{rt}(\vect s)=[I_\text{lr}(\vect s)\: I_\text{sr}(\vect s)]\T$
where $I_\text{road type}(\vect s)$ is an indicator equal to one when $\vect s$ lies on the corresponding road type and zero otherwise. Here, $\text{lr}$ and $\text{sr}$ denote larger road'' and smaller road'', respectively, with ``motorway'' as the reference category.



The models differ on the nature of $u^r(\cdot)$. As before, we indicate with \textsc{WMF} the model where $u^r(\cdot)$ is the weak solution to \eqref{eq:spde_general} with spatially constant parameters, and with \textsc{C-GWMF} and \textsc{B-GWMF} the models where $u^r(\cdot)$ is the solution to \eqref{eq:spde_general} with covariance parameters that are allowed to vary in space.
The \textsc{C-GWMF} model uses the first half of the data to construct a covariate used to describe the covariance structure when fitting the model to the second half of the data. This procedure is the same as in the simulation study. The \textsc{WMF} and \textsc{B-GWMF} models use all data to estimate the field and covariance structure simultaneously.

Two meshes are constructed: a finer one, with maximum spacing $h=0.3$ \si{\kilo\meter} and $\nfine=294$ mesh vertices used to discretize the process $u(\cdot)$, and a second, coarser one, with maximum spacing $\tilde h=0.75$ and $\ncoarse=274$ mesh vertices, used to describe the covariance parameters in the \textsc{B-GWMF} model.

As in Section~\ref{sec:simstudy}, we  use a geometry-influenced prior for $\vect\theta$ for the \textsc{WMF} model. 
The mean for the range parameter is set to 30\% of the diagonal of the domain (width $\times$ height of graph area) while for the marginal standard deviation the mean is 1. The precision for range and marginal standard deviation is $0.1$. 

When we consider the \textsc{C-GWMF} model, we use the transformed median estimates from \textsc{WMF}. For $\log\kappa(\cdot)$ and $\log\tau(\cdot)$ we set means for the intercepts, $b^\kappa_1$ and $b^\tau_1$, by using the transformed median estimates, and we set a zero mean for the coefficients for the covariate $z(\cdot)$, namely $b^\kappa_2$ and $b^\tau_2$. The precisions for all $b^\kappa_j$ $j=1,2$ and $b^\tau_j$ $j=1,2$ are set to 1.

For the prior for $\vect \theta$ when the model is \textsc{B-GWMF}, we continue with the same precision matrix stated in \eqref{eq:mod_Q}, but with parameters $\tau_1=40$ and $\tau_2=1$ and new matrices $\matr C_{\tilde h}$ and $\matr G_{\tilde h}$. The tuning parameters are modified from the previous section to obtain appropriate priors for the new graph considered in this section. As before, the values are chosen from simulating independent realizations from the prior and computing 95\%-quantiles across independent replicates. We compared the prior quantiles to the empirical standard deviations of the data on station-level. We made sure that the prior was flexible enough to capture the observed standard deviation, while penalizing too complex models. The prior is therefore mainly set to accommodate variation in $\sigma(\cdot)$, while we use the same prior for $\rho(\cdot)$, as doing diagnostics for the variation of this field using the data is less obvious. The mean for each $b^\rho_j$ is set to the median for the posterior of $\log\rho_S$ from our fit with \textsc{WMF} for all $j=1,\ldots,\ncoarse$, and similarly for $b^\sigma_j$ we use the median of the posterior for $\log\sigma_S$.

To compare the three models' fit, we consider in-sample prediction scores for each station in the second half of the dataset, which is the same half that \textsc{C-GWMF} uses to fit the full model, that is; $\bigcup_{r=\lceil R/2 \rceil}^R \mathcal S^r$. This is only reported for diagnostic purposes.

To further evaluate the models, we perform LOOCV using the conditional distribution of the kept-out observation conditioned on knowing the other observations and estimated model parameters. 
Let $\vect y^r_{*}=[y^r_{*}(\vect s^r_1)\:\ldots \:y^r_{*}(\vect s^r_{K_r}) ]\T$ be the collection of all observations in locations in $\mathcal S^r_{*}$ for replicate $r$, and we repeat the process for all $r=1,\ldots,R_{*}$. 
Then, we know that 
$$
\vect y^r_{*}\mid \vect\beta,\vect \theta, \sigma_\mathrm{N}\sim \mathcal N_{K_r}({{\matr X}^r_{*}}\T\vect\beta, ({\matr Q^r_{*}})^{-1}),
$$ 
where ${\matr X}^r_{*}$ is the $K_r \times p$ matrix containing the covariate evaluated in all $\mathcal S^r_{*}$, 
$$
{\matr Q^r_{*}}= (\matr A^r_{h,*} \matr Q_{h,q}(\vect \theta)^{-1} {\matr A^r_{h,*}}\T + \sigma_\mathrm{N}^2 \matr I)^{-1},
$$ 
where $\matr Q_{h,q}(\vect\theta)$ is either $\matr Q_{h,1}(\vect\theta)$ or $\matr Q_{h,2}(\vect\theta)$ depending on the model that was fit, and $\matr A^r_{h,*}$ is a projection matrix for the locations in $\mathcal S^r_{*}$ from the mesh $\mathcal V_h$. Let $q^r_{ij}=[\matr Q^r_*]_{ij}$ and $\vect y^r_{*,-i}=[y^r(\vect s_j)]_{j\ne i}$ denote the collection of all observations in replicate $r$ where $y^r(\vect s_i)$ is kept out. 
We obtain the conditional distribution $y^r_i \mid \vect y^r_{*,-i}, \vect\beta=\hat{\vect\beta},\vect\theta=\hat{\vect\theta}, \sigma_\mathrm{N}=\widehat{\sigma_\mathrm{N}}$ using the property of the Gaussian distribution \citep[Chapter 2, Section 2.8]{Gelfand2010}. Our prediction is then associated with the mean 
$$
\mu^r_{\vect s_i} = \hat{\vect\beta} \T \vect x(\vect s_i) - 1/ q^r_{ii} \sum_{j \ne i}  q^r_{ij} \left(y^r(\vect s_j) - \hat{\vect\beta} \T \vect x(\vect s_j)\right)
$$ 
and uncertainty 
$$
\sigma^r_{\vect s_i} = \sqrt{1/ q^r_{ii}}.
$$ 
Note that this prediction is assumes $\vect \beta$, $\vect\theta$ and $\sigma_\mathrm{N}$ are fixed, and we do the same empirical Bayes prediction (no uncertainty in the model parameters is considered in this held-out-replicate LOOCV). We apply the same empirical Bayes prediction approach for all three model types.

RMSE, MAE and CRPS for the models' predictions compared to the observed values in those locations are reported. That is, we compute mean RMSE scores for each location $\vect s_k\in \mathcal S^r_{*}$ across replicates $r=1,\ldots,R_{*}$. The formulas are as follows
\begin{equation*}
	\overline{\text{RMSE}}_y=\frac 1 R \sum^R_{r=1} \sqrt{\frac{1}{K_r}\sum_{\vect s_k\in\mathcal S^r_{*}} (y^r(\vect s_k) - \mu^r_{\vect s_k})^2},
\end{equation*}
and mean MAE
\begin{equation*}
	\overline{\text{MAE}}_y=\frac 1 R \sum^R_{r=1} \frac{1}{K_r}\sum_{\vect s_k\in\mathcal S^r_{*}} \lvert y^r(\vect s_k) - \mu^r_{\vect s_k}\rvert,
\end{equation*}
where $y^r(\vect s_k) $ is the observation and $\mu^r_{\vect s_k}$ is the model prediction in location $\vect s_k\in\mathcal S^r_{*}$ for replicate $r=1,\ldots,R$.
Similarly for mean CRPS, we compute
\begin{equation*}
	\overline{\text{CRPS}}_y=\frac 1 R\sum^R_{r=1} \frac{1}{K_r} \sum_{\vect s_k\in\mathcal S^r_{*}} \text{crps}(y^r(\vect s_k), \mu^r_{\vect s_k}, \sigma^r_{\vect s_k}),
\end{equation*}
where $\sigma^r_{\vect s_k}$ is the standard deviation associated with the prediction $\hat y^r(\vect s_k)$, and $\text{crps}(\cdot,\cdot,\cdot)$ is as stated in \eqref{eq:crps_normal}.

\subsection{Results}
We fit \textsc{WMF} to the full dataset, which has a runtime of 5 seconds. We then use the estimated range and marginal standard deviation parameters to initialize \textsc{B-GWMF} and \textsc{C-GWMF}. 
For initialization of \textsc{B-GWMF}, we use the median of $\log\rho_S$ as the start value for all $b_i^\rho$, $i=1,\ldots,\ncoarse$, and $\log\sigma_S$ as the start value for all $b_i^\sigma$, $i=1,\ldots,\ncoarse$.
\textsc{B-GWMF} has a runtime of ~3 hours and 35 minutes. For that reason, any cross-validation approach with repeated model fits for model evaluation is not feasible.
\textsc{C-GWMF} is initialized with prior means for $b_1^\kappa$, $b_1^\tau$, $b_2^\kappa$ and $b_2^\tau$. \textsc{C-GWMF} has a total runtime of 12 seconds, including construction of covariates (9 seconds) and full model fit (3 seconds).
Scoring rules computed for the fitted model, evaluated on the data it was fitted to, are displayed in Table~\ref{tab:madrid_model_comparison}.

\begin{table}[ht]
	\centering
	\caption{Observation-level comparison of fitted log traffic intensity for the Madrid case study. Scores are computed by comparing each observed log intensity with the corresponding replicate-specific fitted value and uncertainty (standard deviation) at the same station, only for the second half of the replicates. Lower values indicate better performance for RMSE, MAE and CRPS. Note that these are ``in-sample'' scores, and cannot be used to evaluate the predictive ability of the models.}
	\label{tab:madrid_model_comparison}
	\begin{tabular}{lrrrr}
		\toprule
		Model & $\sum_{r=\lceil R/2\rceil}^R N_r$ & $\overline{\text{RMSE}}_y$ & $\overline{\text{MAE}}_y$ & $\overline{\text{CRPS}}_y$  \\
		\midrule
		\textsc{B-GWMF} & 735 & 0.0772 & 0.0453 & 0.0461 \\
		\textsc{C-GWMF} & 735 & 0.1379 & 0.1069 & 0.0859 \\
		\textsc{WMF}  & 735 & 0.1039 & 0.0752 & 0.0670 \\
		\bottomrule
	\end{tabular}
\end{table}

Because of the high computational costs, the \textsc{B-GWMF} model is fitted using an empirical Bayes strategy. Consequently, uncertainty in the covariance parameters is not propagated in the same way as for the lower-dimensional \textsc{WMF} and \textsc{C-GWMF} model. We observe that \textsc{B-GWMF} is ``preferred'' with respect to all scores considered in Table \ref{tab:madrid_model_comparison}. Again, we underline that the scores are only meant to illustrate how well the models are fitted to the observed values. Note that the RMSE score for the \textsc{C-GWMF} model is higher than for the two other two. This   can be explained by looking at the estimated noise parameter $\sigma_\mathrm{N}$: this is  0.31 (0.25,0.42) for the \textsc{C-GWMF} model, and  0.23 (0.20,0.26) for the \textsc{WMF} model. For \textsc{B-GWMF}, the estimate is even lower, at 0.15 (0.14, 0.17). Thus, \textsc{B-GWMF} has the lowest estimated noise level of the three models, which is consistent with its lower RMSE.

\begin{table}[ht]
\centering
\caption{Posterior median fixed-effect estimates, with 95\% credible intervals in parentheses. The reference category is motorway.}
\begin{tabular}{lccc}
\toprule
Model & $\beta_{\text{ref}}$ & $\beta_{\text{lr}}$ & $\beta_{\text{sr}}$ \\
\hline
\textsc{B-GWMF}
& $7.56\;(7.45,\;7.68)$ & $-1.38\;(-1.50,\;-1.26)$ & $-2.90\;(-3.04,\;-2.76)$ \\
\textsc{C-GWMF} & $7.55\;(7.36,\;7.74)$ & $-1.34\;(-1.53,\;-1.13)$ & $-2.88\;(-3.11,\;-2.64)$ \\
\textsc{WMF} & $7.59\;(7.46,\;7.71)$ & $-1.41\;(-1.54,\;-1.28)$ & $-2.94\;(-3.09,\;-2.78)$ \\
\bottomrule
\end{tabular}
\label{tab:madrid-mean-effects}
\end{table}

We report the fixed effects estimates for each model in Table \ref{tab:madrid-mean-effects}. Estimates are quite consistent among the three models:   as expected, the log-traffic intensity is the highest for the reference category (highway), and lowest for the smaller road category.

\begin{figure}[htb]
	\centering
	\subfigure[$\hat{\rho}(\vect s)$ (in \si{\kilo\meter})\label{fig:case_range_b-gwmf}]{\includegraphics[width=0.45\linewidth]{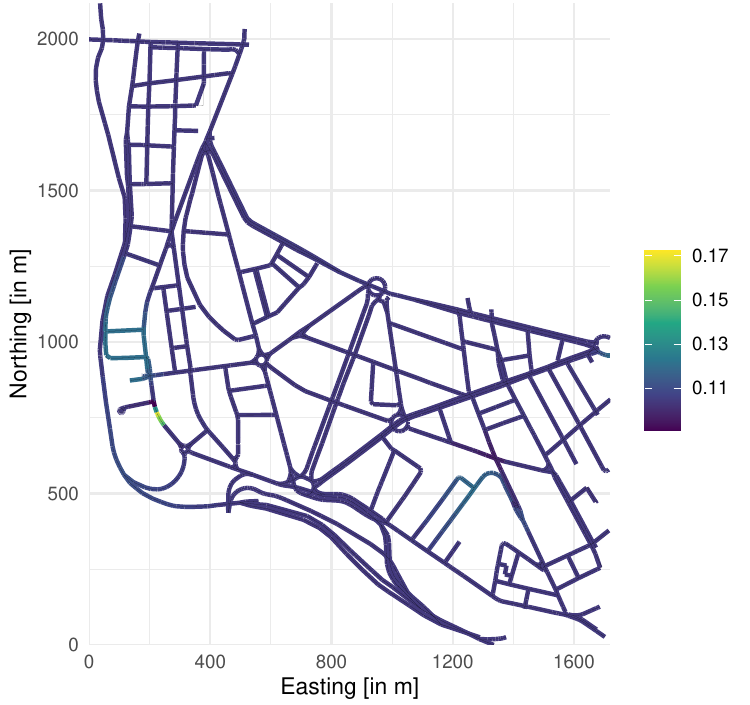}}
	\subfigure[$\hat\sigma(\vect s)$\label{fig:case_sigma_b-gwmf}]{\includegraphics[width=0.45\linewidth]{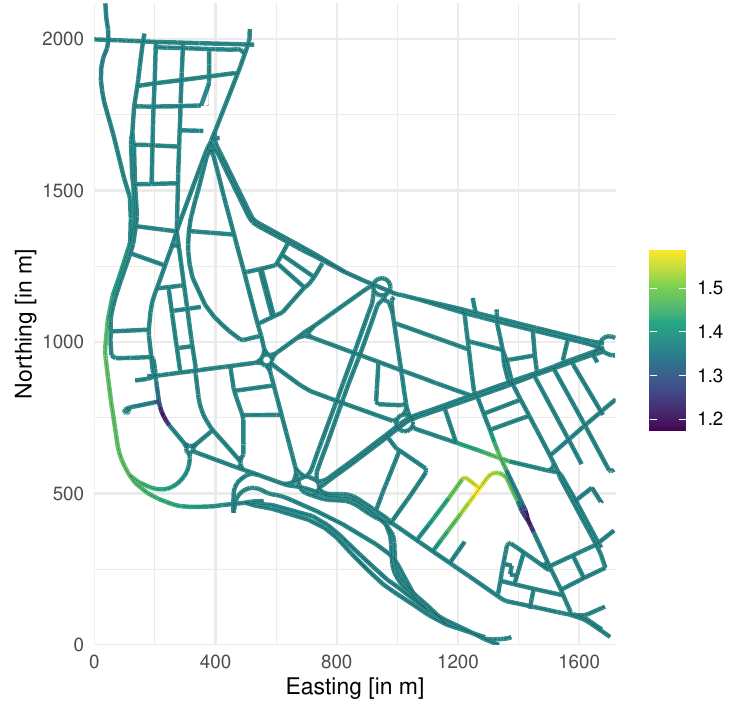}}
	\subfigure[$\hat\rho(\vect s)=2/\hat{\kappa}(\vect s)$ (in \si{\kilo\meter})\label{fig:case_range_c-gwmf}]{\includegraphics[width=0.45\linewidth]{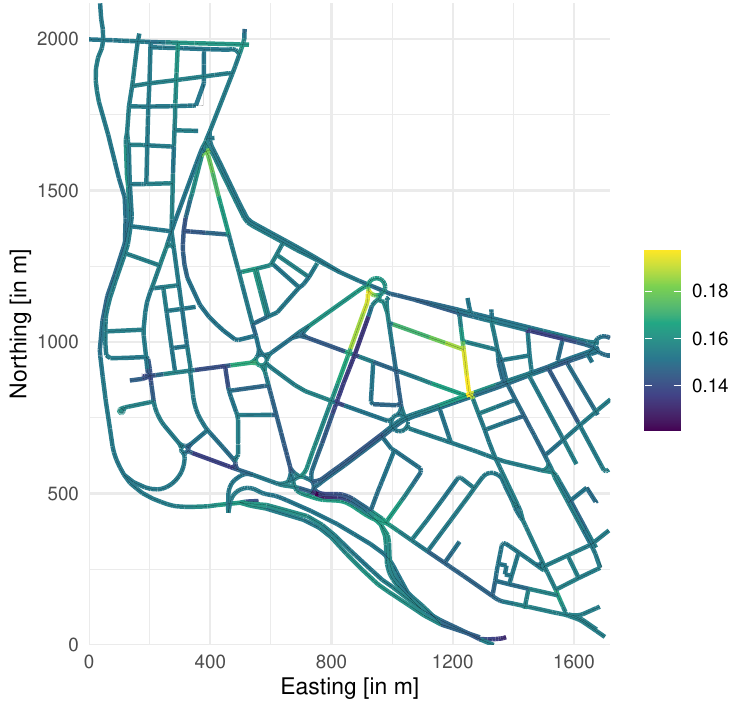}}
	\subfigure[$\hat\sigma(\vect s) =1/({2\hat\kappa(\vect s)^{3/2}\hat\tau(\vect s)})$\label{fig:case_sigma_c-gwmf}]{\includegraphics[width=0.45\linewidth]{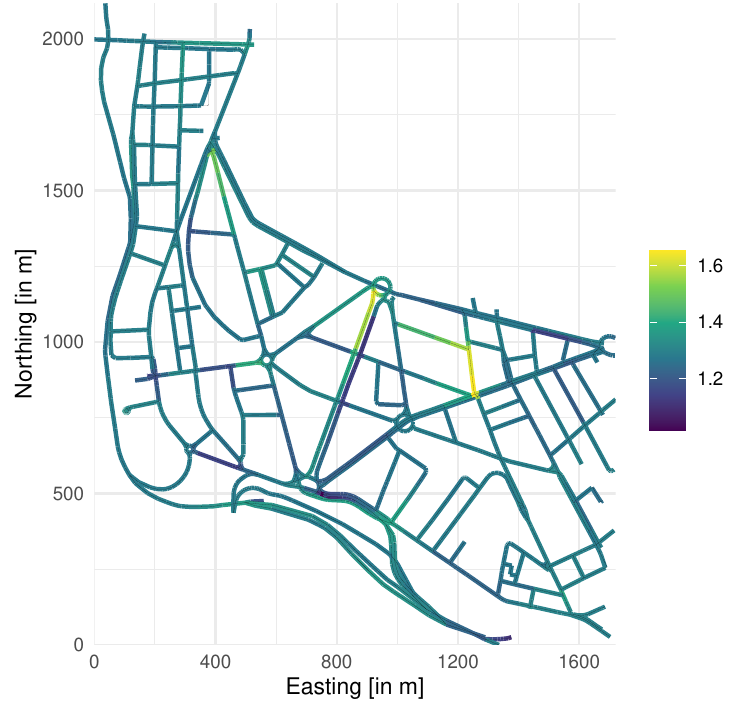}}
	\caption{The covariance structure estimated by \textsc{B-GWMF} for model parameters (a) $\rho(\cdot)$ and (b) $\sigma(\cdot)$ and fields from \textsc{C-GWMF} using estimated fields $\hat\kappa(\cdot)$ and $\hat\tau(\cdot)$ to obtain comparable fields using the transformations in \eqref{eq:param_transformation} for (c) ``range'' and (d) ``marginal standard deviation''.}
	\label{fig:case_est_non_stat_fields}
\end{figure}

Next we look into the estimated covariance structure of each model. The \textsc{WMF} model estimates the  range as 0.10 \si{\kilo\meter} (95\% CI: (0.09,0.12)). The space-varying covariance parameters as estimated  using the \textsc{B-GWMF} and \textsc{C-GWMF} models are shown in Figure~\ref{fig:case_est_non_stat_fields}. For the \textsc{B-GWMF}, the range 
(Figure~\ref{fig:case_range_b-gwmf}) is mostly around 0.11 \si{\kilo\meter}, but it gets higher in some areas. 
\textsc{C-GWMF} also estimates some spatial variation in the transformed field $\hat\rho(\cdot)=2/\hat\kappa(\cdot)$, where the field is mainly centered around 0.16 \si{\kilo\meter} and has some local increases up to above 0.18
\si{\kilo\meter} (which is longer than the ``range'' estimated by \textsc{B-GWMF}, and in different regions). 

For the marginal standard deviation, the \textsc{WMF} estimate is 1.37 (95\% CI: (1.29,1.46)). The \textsc{B-GWMF} model estimated the standard deviation, varying in space from 1.2 to 1.6, as shown in Figure~\ref{fig:case_sigma_b-gwmf}. 
Also the \textsc{C-GWMF} model, estimated the marginal variance field (Figure~\ref{fig:case_sigma_c-gwmf}) to be centered around 1.3, with local increases to above 1.6 and local decreases down to 1.0. Note that the pattern on the estimated standard deviation is, as expected, similar to the one of the constructed covariate in Figure~\ref{fig:logsd_for_cgwmf}, i.e.\ the covariate is scaled with a positive coefficient.

We recall the model parameter for observation noise, $\sigma_\mathrm{N}$, and that \textsc{B-GWMF} has the lowest estimated noise. That is, it is estimating little variation from the observed values, and we can suspect over-fitting.
To further check of over-fitting, we test the three models' predictive power on a new dataset, comparing scores of predictions using conditional distributions of the left-out observations conditioned on the other observations and the estimated parameters.

\begin{table}[ht]
	\centering
	\caption{Conditional leave-one-out scores on the January--February 2026 Madrid data, using parameters fitted to the 2023--2025 data. Each score is an based on marginal predictions of one station conditional on the remaining stations from the same replicate.}
	\label{tab:madrid-loocv-2026}
	\begin{tabular}{lrrrr}
		\toprule
		Model & $\sum_{r=1}^{R_{*}} K_r$ & $\overline{\text{RMSE}}_y$ & $\overline{\text{MAE}}_y$ & $\overline{\text{CRPS}}_y$ \\
		\midrule
		\textsc{B-GWMF} & 389 & 0.824 & 0.637 & 0.466 \\
		\textsc{C-GWMF} & 389 & 0.812 & 0.625 & 0.451 \\
		\textsc{WMF}  & 389 &  0.802 & 0.620 & 0.447 \\
		\bottomrule
	\end{tabular}
\end{table}

The out-of-sample comparison reveals that \textsc{B-GWMF} performs worse on the new replicates with respect to all scores in Table~\ref{tab:madrid-loocv-2026}. Second best model is \textsc{C-GWMF} and \textsc{WMF} is the model with the lowest scores. 
Since the two GWMF-based models have starting values equal to the \textsc{WMF} model, and are allowed to explore more complex structures to obtain a better fit - but then end up performing worse on the test dataset, we cannot rule out overfitting for the more complex models in this case.

\section{Discussion}
\label{sec:discussion}
Motivated by traffic modeling on a road network, we have introduced a new GRF model on metric graphs with flexible covariance structure controlled by spatial basis functions. 
The main goal was to investigate when models with spatially varying covariance parameters are useful, and how they compare with simpler models with spatially constant covariance parameters. We consider three approaches: a Whittle--Mat\'ern field (\textsc{WMF}) with spatially constant coefficients, a model with spatially-varying coefficients based on a covariate constructed from splitting the data (\textsc{C-GWMF}), and a basis-function representation of the covariance fields (\textsc{B-GWMF}). The latter is the novel approach and is inspired by the methods of \citet{Fuglstad2015nonstat}. 

In the simulation study, we find that when the truth is a GWM-GRF with spatially varying covariance coefficients and we consider spatial prediction in unobserved locations and identification of the true covariance structure, the two GWM-GRFs, \textsc{C-GWMF} and \textsc{B-GWMF}, can outperform the WM-GRF, \textsc{WMF}. Additionally, we find that in the setting where the true covariance parameters are constant in space, all three models perform similarly under enough spatial and temporal data coverage, but there is a preference for \textsc{WMF} in this simple scenario.

In a setting where the data coverage is low, i.e., the number of replicates is low \textit{and} spatial coverage is low, \textsc{WMF} is the most reliable choice, as the other two methods rely on enough replicates and spatial coverage. \textsc{WMF} is computationally efficient, and performs similarly to the other models in the scenario with lowest temporal and spatial coverage in the simulation study.
When temporal coverage is sufficient, and a spatially varying  covariance structure is evident, we find that the two approaches considered are outperforming the simplest model with respect to spatial prediction and recognition of the true covariance structure. The spatially varying covariance structure is most accurately identified with \textsc{B-GWMF}.
The simulation study also investigates robustness to different noise levels. When the measurement standard deviation was increased by a factor 10, the results show the same trends presented in Section~\ref{sec:simstudy}, while factor 50, we see that the model choice is less important to obtain accurate predictions and identify the covariance structure.

Traffic data is well suited for the flexible covariance models we consider in this study since it is collected with extensive temporal resolution. We considered an application of traffic counts in one small area of a specific district of Madrid with 62 measurement stations.
We find that the flexible model using basis functions can obtain competitive in-sample scores.
However, the the out-of-sample LOOCV scores do not show the same advantage for the flexible model as in the simulation study. This suggests that the flexibility of B-GWMF can lead to overfitting or underestimated predictive uncertainty when the data are less informative than in the idealized simulation setting.

Prior specification of basis function parameters used an approach where spatially neighboring parameters are shrunk towards each other through a WM prior. More careful choice of the penalization parameter, or other approaches could be explored to improve prediction with \textsc{B-GWMF} beyond the current implementation. Inference is not fully Bayesian, as the full marginals for the basis function parameters are not explored. Further exploration of methods to do full Bayesian analysis in a high-dimensional space is of interest to make \textsc{B-GWMF} comparable to the fully Bayesian methods that include \textsc{WMF} and \textsc{C-GWMF}.

The flexible covariance structure in \textsc{B-GWMF} is limited to a modestly-sized graph due to the computational complexity. We found that the graph considered in the application is close to the practical limit in the current implementation of \texttt{R-INLA}. The computational time was additionally heavily influenced by the prior restriction/penalization on $\vect \theta$. By increasing the dependency between neighboring coefficients, one can reduce the computation time drastically. 
Alternatively, one could look into setting up the optimization outside of \texttt{R-INLA}, to make it more problem-specific, as \textsc{B-GWMF} is outside of the typical models that \texttt{R-INLA} is meant for (typically the hyperparameter space should not be too large). In this paper we used \texttt{R-INLA} for all three models for convenience of fitting complex models using the \texttt{MetricGraph}- and \texttt{rSPDE}-packages. 

We believe that applying \textsc{B-GWMF} to other spatial domains, or other data sources, where identification of the underlying covariance structures is of interest is a natural next step from the work presented. While temporal and/or spatial coverage might be limiting factors in many real-life applications, the possible improvements in spatial prediction and recognition of the true covariance structures with our flexible model are apparent from the results in the simulation study.
Overall, the results suggest that flexible covariance structures on metric graphs can be valuable when the data contain enough spatial and replicated information to identify the additional structure, but that less flexible models remain important baselines due to their stability, interpretability, and substantially lower computational cost. 

\section*{Acknowledgments}
This research was funded by The Research Council of Norway's IKTPLUSS program, project number 332237. 

\section*{Declaration of generative AI and AI-assisted technologies in the manuscript preparation process}
During the preparation of this work the authors used ChatGPT 5.5/5.6 in order to track progression of the project, as a coding assistant and grammar checks of the manuscript. After using this tool/service, the authors reviewed and edited the content as needed and take full responsibility for the content of the published article.

\bibliography{references}  

\appendix{
	\input{appendix_A}
	\input{appendix_B}

}

\end{document}

%% file: appendix_A.tex
\section{Unique solution of the SPDE}\label{app:A}
Inspired by the SPDE approach, we investigate the possibility of using a basis for the solution $f$ of the given SDE
\begin{equation}\label{eq:spde_f_appA}
    (\gamma^2 - \Delta)(f) = \mathcal W
\end{equation}
where the solution $f$ can be expressed as a sum of some basis $\{\varphi_k\}_{k=1}^\infty$,
\begin{equation}\label{eq:basis_f}
    f(t) = \sum^\infty_{k=0}f_k\varphi_k(t).
\end{equation}
We decide the basis functions from the eigenfunctions of the Laplace operator with Neumann boundary conditions on end points $t=0,1$,
\begin{equation}
    \varphi_0(t)=1 \quad\quad \varphi_k(t) = \sqrt 2 \cos{(k\pi t)}\quad k=1,2\ldots
\end{equation}
where the corresponding eigenvalues are $\lambda_k=(k\pi)^2$. The Gaussian white noise process can similarly be expanded,
\begin{equation}\label{eq:basis_noise}
    \mathcal W(t) = \sum_{k=0}^\infty \xi_k\varphi_k(t)
\end{equation}
where $\xi_k$ are i.i.d. $N(0,1)$. 
By inserting \eqref{eq:basis_f} and \eqref{eq:basis_noise} into \eqref{eq:spde_f_appA} we obtain 
\begin{equation*}
    (\gamma^2 - \Delta)(f) = \gamma^2\sum^\infty_{k=0}f_k\varphi_k(t) + \sum^\infty_{k=0}f_k\lambda_k\varphi_k(t) = \sum_{k=0}^\infty \xi_k\varphi_k(t)
\end{equation*}
where we only need to compare the random variables pairwise for each basis function $\varphi_k()$, and we find
\begin{equation}
    f_k = \frac{\xi_k}{\gamma^2 + \lambda_k} =  \frac{\xi_k}{\gamma^2 + (k\pi)^2} \sim N\left(0,1/(\gamma^2+(k\pi)^2)^2\right)\quad k=0,1,\ldots
\end{equation}
as the random coefficients related to the chosen basis. In particular we note that $k=0$ gives us $f_0 = \xi_0/\gamma^2$, where $\gamma$ becomes small (tends to zero) would mean that $f(t)$ tend to infinity for any $t$. 
To avoid this, we infer 
\begin{equation}
    \int f(t) dt =  \int f_0\varphi_0(t) dt +  \sum_{k=1}^\infty\int f_k\varphi_k(t) dt = 0
\end{equation}
where the second term will always be zero, so the restriction simplifies to 
\begin{equation}
    \int f_0\varphi_0(t)dt = 0
\end{equation}
which forces $f_0=\xi_0/\gamma^2=0$, i.e. the randomness of $\xi_0$ (zero frequency) is lost.

%% file: appendix_B.tex
\section{Recovery of WMF}\label{app:B}
\begin{figure}[htb]
    \centering
    \subfigure[$\text{RMSE}_\eta$]{\includegraphics[width=0.45\linewidth]{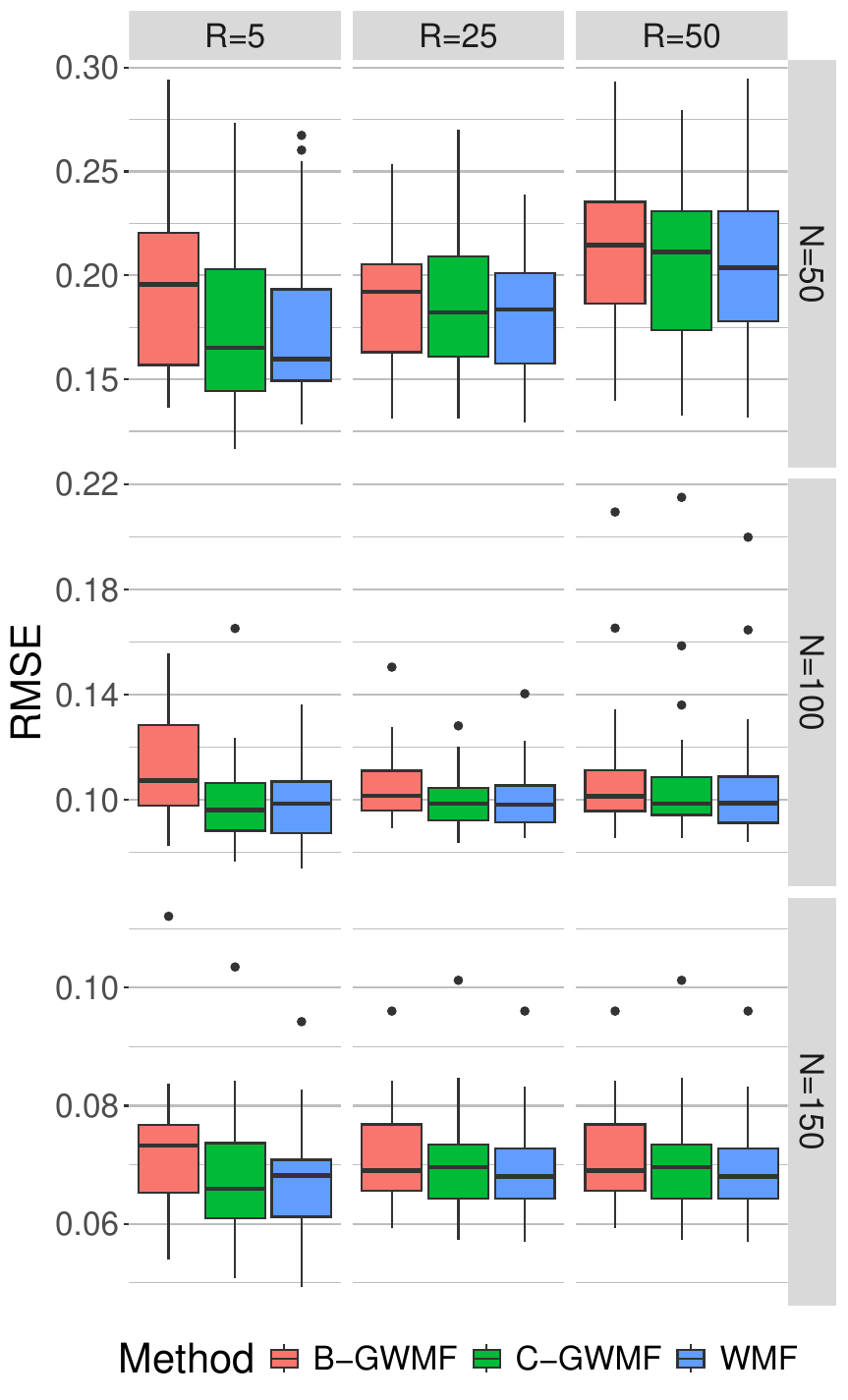}}
    \subfigure[$\text{CRPS}_\eta$]{\includegraphics[width=0.45\linewidth]{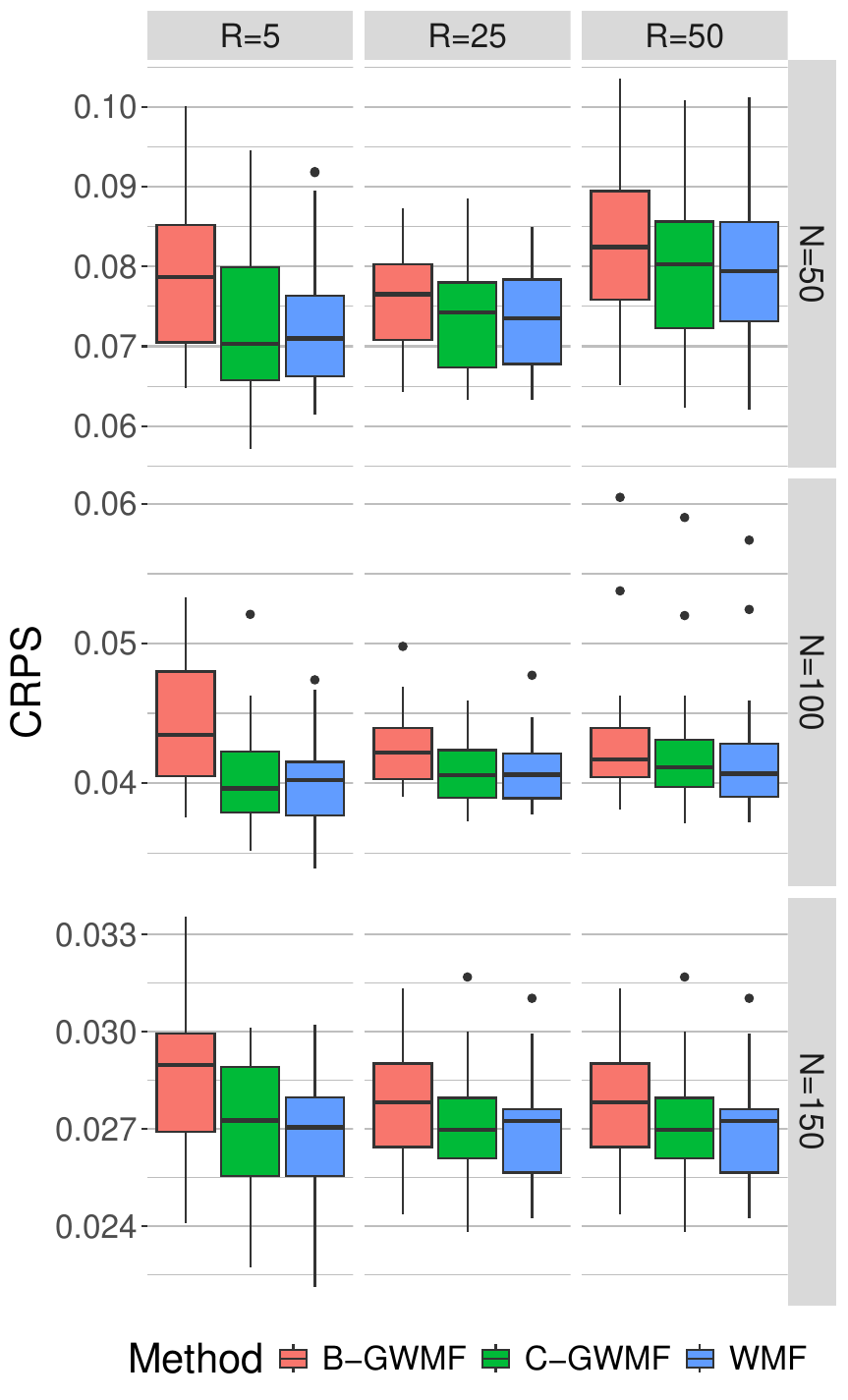}}
    \caption{RMSE and CRPS for the latent field in the scenario when the true covariance structure is stationary. For better visualization, one point was removed for \textsc{B-GWMF} in the setting $N=50$ and $R=50$.}
    \label{fig:rmse_crps_stat}
\end{figure}

\begin{figure}[htb]
    \centering
    \subfigure[RMSE$_{\log\rho}$]{\includegraphics[width=0.45\linewidth]{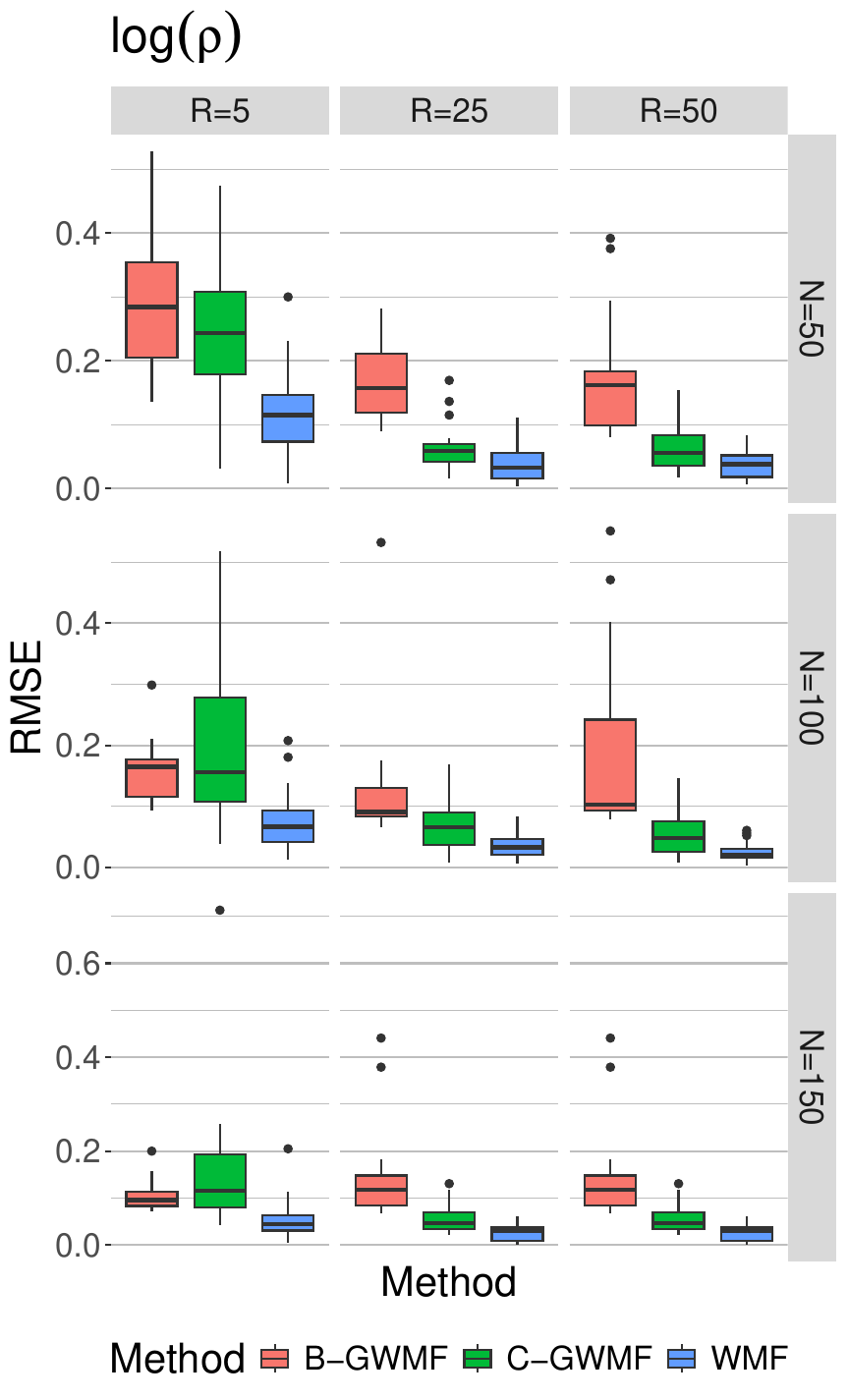}}
    \subfigure[RMSE$_{\log\sigma}$]{\includegraphics[width=0.45\linewidth]{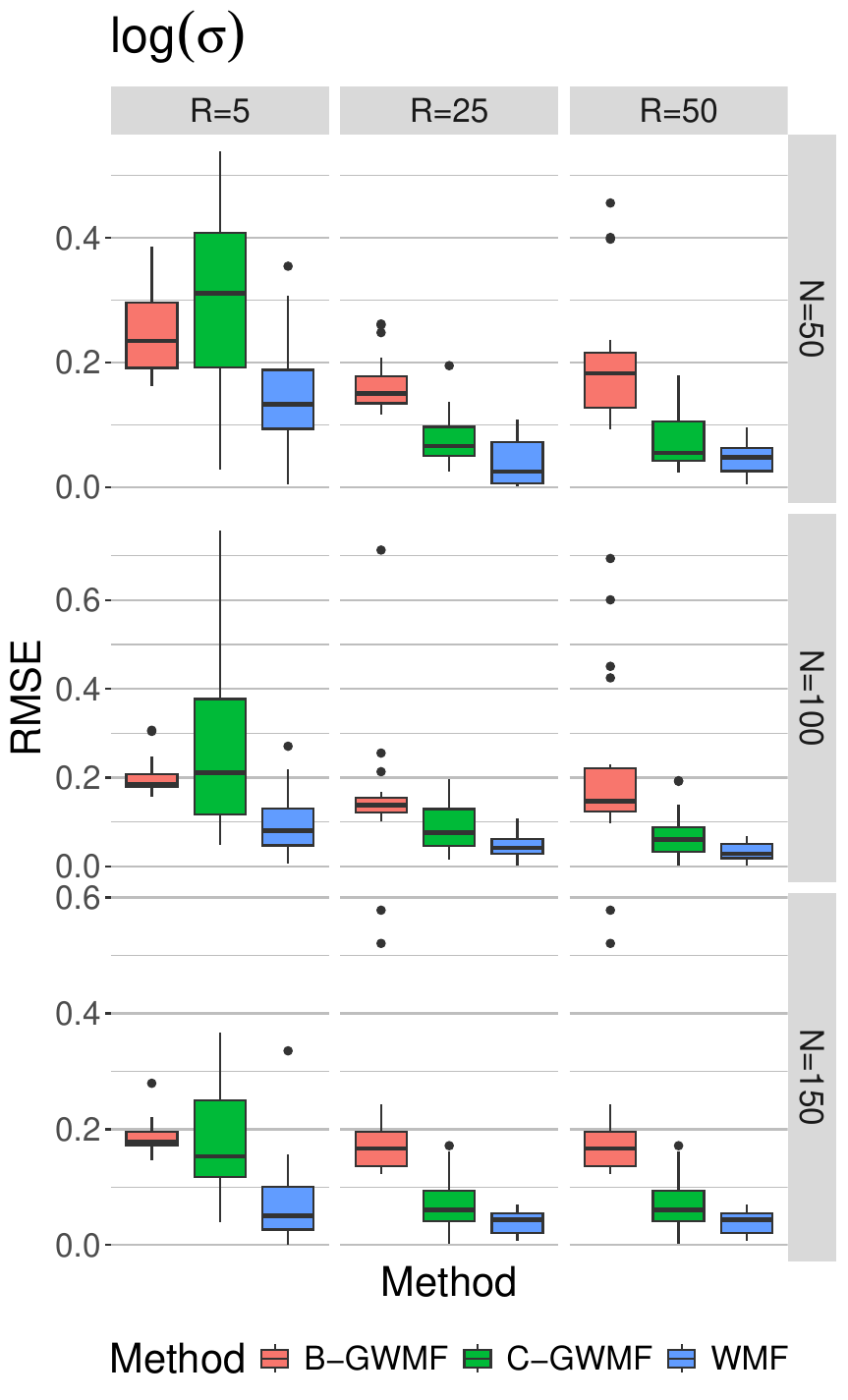}}
    \caption{RMSE for the (a) log-transformed covariance field describing $\log\rho(\cdot)$ and (b) log-transformed covariance field describing $\log\sigma()$ in the scenario when the true covariance structure is stationary. For better visualization, four points were removed; three for \textsc{B-GWMF} in the settings $N=100,150$ and $R=50$, and $N=150$ and $R=25$, and one for \textsc{C-GWMF} in the setting $N=100$ and $R=5$.}
    \label{fig:rmse_range_sdev_stat}
\end{figure}

The results from the simulation study discussed in Section \ref{sec:simstudy} in the scenario when the true covariance parameters $\rho(\cdot)$ and $\sigma(\cdot)$ are invariant to spatial location. Figure~\ref{fig:rmse_crps_stat} presents the average RMSE and CRPS for the latent field prediction and Figure~\ref{fig:rmse_range_sdev_stat} show the RMSE scores for the range parameter $\rho(\cdot)$ and $\sigma(\cdot)$. Note that in this scenario, \textsc{WMF} is the ``true'' model, while \textsc{B-GWMF} and \textsc{C-GWMF} are models that allow for more complex covariance structures.

%% file: references.bib
@book{Diggle2007geostat,
title = "Model-based Geostatistics.",
author = "Diggle, {Peter J.} and Ribeiro, {Paulo J.}",
year = "2007",
month = mar,
language = "English",
isbn = "0387329072 978-0387329079",
series = "Springer Series in Statistics",
publisher = "Springer",
}

@book{Gelfand2010,
	title = {Handbook of Spatial Statistics},
	editor = {Gelfand, A.E. and Diggle, P. and Guttorp, P. and Fuentes, M.},
	ISBN = {9780429136504},
	url = {http://dx.doi.org/10.1201/9781420072884},
	DOI = {10.1201/9781420072884},
	publisher = {CRC Press},
	year = {2010},
	month = Mar 
}

@article{Lindgren2011SPDE,
  title = {{An Explicit Link between Gaussian Fields and Gaussian Markov Random Fields: The Stochastic Partial Differential Equation Approach}},
  volume = {73},
  ISSN = {1467-9868},
  DOI = {10.1111/j.1467-9868.2011.00777.x},
  number = {4},
  journal = {Journal of the Royal Statistical Society Series B: Statistical Methodology},
  publisher = {Oxford University Press (OUP)},
  author = {Lindgren,  Finn and Rue,  Håvard and Lindstr\"{o}m,  Johan},
  year = {2011},
  month = aug,
  pages = {423–498}
}

@article{Rue2009INLA,
    author = {Rue, Håvard and Martino, Sara and Chopin, Nicolas},
    title = {Approximate {B}ayesian inference for latent {G}aussian models by using integrated nested {L}aplace approximations},
    journal = {Journal of the Royal Statistical Society: Series B (Statistical Methodology)},
    volume = {71},
    number = {2},
    pages = {319-392},
    doi = {10.1111/j.1467-9868.2008.00700.x},
    year = {2009}
}

@misc{Lindgren2024inlabru,
      title={inlabru: software for fitting latent {G}aussian models with non-linear predictors}, 
      author={Finn Lindgren and Fabian Bachl and Janine Illian and Man Ho Suen and Håvard Rue and Andrew E. Seaton},
      year={2024},
      eprint={2407.00791},
      archivePrefix={arXiv},
      primaryClass={stat.ME},
}

@Manual{Bolin2019rSPDE,
  title = {{rSPDE: Rational Approximations of Fractional Stochastic
      Partial Differential Equations}},
  DOI = {10.32614/cran.package.rspde},
  journal = {CRAN: Contributed Packages},
  publisher = {The R Foundation},
  author = {Bolin,  David and Simas,  Alexandre},
  year = {2019},
  note = {{R} package version 2.3.3},
  month = aug 
}

@misc{Bolin2023statinfmetricgraph,
      title={Statistical inference for {G}aussian {W}hittle-{M}at\'ern fields on metric graphs}, 
      author={David Bolin and Alexandre Simas and Jonas Wallin},
      year={2023},
      eprint={2304.10372},
      archivePrefix={arXiv},
      primaryClass={stat.ME},
}

@article{Bolin2024GWMmetricgraph,
  title={Gaussian {W}hittle--{M}at{\'e}rn fields on metric graphs},
  author={Bolin, David and Simas, Alexandre B and Wallin, Jonas},
  journal={Bernoulli},
  volume={30},
  number={2},
  DOI = {10.3150/23-bej1647},
  pages={1611--1639},
  year={2024},
  publisher={Bernoulli Society for Mathematical Statistics and Probability}
}

@misc{Bolin2025nonstat,
      title={A new class of non-stationary {G}aussian fields with general smoothness on metric graphs}, 
      author={David Bolin and Lenin Riera-Segura and Alexandre B. Simas},
      year={2025},
      eprint={2501.11738},
      archivePrefix={arXiv},
      primaryClass={stat.ME},
}

@misc{Bolin2025loggausscox,
      title={{Log-Gaussian Cox Processes on General Metric Graphs}}, 
      author={David Bolin and Damilya Saduakhas and Alexandre B. Simas},
      year={2025},
      eprint={2501.18558},
      archivePrefix={arXiv},
      primaryClass={stat.ME}, 
}

@Manual{Bolin2023MetricGraph,
  title = {MetricGraph: Random fields on metric graphs},
  author = {David Bolin and Alexandre B. Simas and Jonas Wallin},
  year = {2023},
  note = {{R} package version 1.3.0.9000},
  doi = {10.32614/CRAN.package.MetricGraph}
}

@Manual{osm,      
	author = {{OpenStreetMap contributors}},
    title = {{Planet dump retrieved from https://planet.osm.org }},
    howpublished = "\url{ https://www.openstreetmap.org}",
    year = {2017},
    note = {Data extraction: 2025-01-17}
}

@Article{cran-osmdata,
	title = {osmdata},
	author = {{Mark Padgham} and {Bob Rudis} and {Robin Lovelace} and
	{Maëlle Salmon}},
	journal = {Journal of Open Source Software},
	year = {2017},
	volume = {2},
	number = {14},
	pages = {305},
	month = {jun},
	publisher = {The Open Journal},
	url = {https://joss.theoj.org/papers/10.21105/joss.00305},
	doi = {10.21105/joss.00305},
}

@Book{sf1,
    author = {Edzer Pebesma and Roger Bivand},
    title = {{Spatial Data Science: With applications in R}},
    year = {2023},
    publisher = {{Chapman and Hall/CRC}},
    doi = {10.1201/9780429459016},
  }

@Article{sf,
    author = {Edzer Pebesma},
    title = {{Simple Features for R: Standardized Support for Spatial
      Vector Data}},
    year = {2018},
    journal = {{The R Journal}},
    doi = {10.32614/RJ-2018-009},
    pages = {439--446},
    volume = {10},
    number = {1},
}

@article{Anderes2020isotropic,
  title = {Isotropic covariance functions on graphs and their edges},
  volume = {48},
  ISSN = {0090-5364},
  DOI = {10.1214/19-aos1896},
  number = {4},
  journal = {The Annals of Statistics},
  publisher = {Institute of Mathematical Statistics},
  author = {Anderes,  Ethan and Møller,  Jesper and Rasmussen,  Jakob G.},
  year = {2020},
  month = aug 
}

@article{Hoef2006spatial,
  title={Spatial statistical models that use flow and stream distance},
  author={Hoef, Jay M Ver and Peterson, Erin and Theobald, David},
  journal={Environmental and Ecological statistics},
  volume={13},
  doi = {10.1007/s10651-006-0022-8},
  pages={449--464},
  year={2006},
  publisher={Springer}
}

@article{Hoef2010moving,
  title={A moving average approach for spatial statistical models of stream networks},
  author={Ver Hoef, Jay M and Peterson, Erin E},
  journal={Journal of the American Statistical Association},
  volume={105},
  DOI = {10.1198/jasa.2009.ap08248},
  number={489},
  pages={6--18},
  year={2010},
  publisher={Taylor \& Francis}
}

@article{Ingebrigtsen2015,
title = {Estimation of a non-stationary model for annual precipitation in southern {N}orway using replicates of the spatial field},
journal = {Spatial Statistics},
volume = {14},
pages = {338-364},
year = {2015},
issn = {2211-6753},
doi = {10.1016/j.spasta.2015.07.003},
author = {Rikke Ingebrigtsen and Finn Lindgren and Ingelin Steinsland and Sara Martino}
}

@article{Fuglstad2015nonstat,
 ISSN = {10170405, 19968507},
 URL = {http://www.jstor.org/stable/24311007},
 author = {Geir-Arne Fuglstad and Finn Lindgren and Daniel Simpson and Håvard Rue},
 journal = {Statistica Sinica},
 number = {1},
 pages = {115--133},
 publisher = {Institute of Statistical Science, Academia Sinica},
 title = {Exploring a new class of non-stationary spatial Gaussian random fields with varying local anisotropy},
 urldate = {2026-02-06},
 volume = {25},
 year = {2015}
}

@article{Fuglstad2015doesnonstat,
  title = {Does non-stationary spatial data always require non-stationary random fields?},
  volume = {14},
  ISSN = {2211-6753},
  url = {http://dx.doi.org/10.1016/j.spasta.2015.10.001},
  DOI = {10.1016/j.spasta.2015.10.001},
  journal = {Spatial Statistics},
  publisher = {Elsevier BV},
  author = {Fuglstad,  Geir-Arne and Simpson,  Daniel and Lindgren,  Finn and Rue,  Håvard},
  year = {2015},
  month = nov,
  pages = {505–531}
}

@article{Lilleborge2026,
	title = {Joint modeling of line and point data on metric graphs},
	volume = {71},
	ISSN = {2211-6753},
	url = {http://dx.doi.org/10.1016/j.spasta.2025.100946},
	DOI = {10.1016/j.spasta.2025.100946},
	journal = {Spatial Statistics},
	publisher = {Elsevier BV},
	author = {Lilleborge,  Karina and Martino,  Sara and Fuglstad,  Geir-Arne and Lindgren,  Finn and Ingebrigtsen,  Rikke},
	year = {2026},
	month = Mar,
	pages = {100946}
}

@book{banerjee2026hierarchical,
  title={Hierarchical modeling and analysis for spatial data},
  author={Banerjee, Sudipto and Gelfand, Alan E and Carlin, Bradley P},
  year={2026},
  publisher={CRC Press},
  address = {FL: Boca Roten}
}

@InProceedings{Borovitskiy2021MaternOnGraphs,
  title = 	 { Mat{é}rn Gaussian Processes on Graphs },
  author =       {Borovitskiy, Viacheslav and Azangulov, Iskander and Terenin, Alexander and Mostowsky, Peter and Deisenroth, Marc and Durrande, Nicolas},
  booktitle = 	 {Proceedings of The 24th International Conference on Artificial Intelligence and Statistics},
  pages = 	 {2593--2601},
  year = 	 {2021},
  editor = 	 {Banerjee, Arindam and Fukumizu, Kenji},
  volume = 	 {130},
  series = 	 {Proceedings of Machine Learning Research},
  month = 	 {13--15 Apr},
  publisher =    {PMLR},
  url = 	 {https://proceedings.mlr.press/v130/borovitskiy21a.html}
}

@article{SanzAlonso2022SPDEGraph,
  title = {The SPDE Approach to Matérn Fields: Graph Representations},
  volume = {37},
  ISSN = {0883-4237},
  url = {http://dx.doi.org/10.1214/21-sts838},
  DOI = {10.1214/21-sts838},
  number = {4},
  journal = {Statistical Science},
  publisher = {Institute of Mathematical Statistics},
  author = {Sanz-Alonso,  Daniel and Yang,  Ruiyi},
  year = {2022},
  month = Nov 
}

@article{LOWRY201498,
title = {Spatial interpolation of traffic counts based on origin–destination centrality},
journal = {Journal of Transport Geography},
volume = {36},
pages = {98-105},
year = {2014},
issn = {0966-6923},
doi = {https://doi.org/10.1016/j.jtrangeo.2014.03.007},
url = {https://www.sciencedirect.com/science/article/pii/S0966692314000519},
author = {Michael Lowry}}

@article{PULUGURTHA2021103071,
title = {Modeling AADT on local functionally classified roads using land use, road density, and nearest nonlocal road data},
journal = {Journal of Transport Geography},
volume = {93},
pages = {103071},
year = {2021},
issn = {0966-6923},
doi = {https://doi.org/10.1016/j.jtrangeo.2021.103071},
url = {https://www.sciencedirect.com/science/article/pii/S0966692321001241},
author = {Srinivas S. Pulugurtha and Sonu Mathew}}

@article{BarringtonLeigh2017,
  author  = {Barrington-Leigh, Christopher and Millard-Ball, Adam},
  title   = {The world's user-generated road map is more than 80\% complete},
  journal = {PLOS ONE},
  year    = {2017},
  volume  = {12},
  number  = {8},
  pages   = {e0180698},
  doi     = {10.1371/journal.pone.0180698}
}

@article{Haklay2010,
  author  = {Haklay, Mordechai},
  title   = {How Good is Volunteered Geographical Information? A Comparative Study of OpenStreetMap and Ordnance Survey Datasets},
  journal = {Environment and Planning B: Planning and Design},
  year    = {2010},
  volume  = {37},
  number  = {4},
  pages   = {682--703},
  doi     = {10.1068/b35097}
}
